\documentclass[sigconf, screen]{acmart}
\pdfoutput=1

\AtBeginDocument{%
  \providecommand\BibTeX{{%
    \normalfont B\kern-0.5em{\scshape i\kern-0.25em b}\kern-0.8em\TeX}}}

\copyrightyear{2021}
\acmYear{2021}
\setcopyright{acmcopyright}\acmConference[CHI '21]{CHI Conference on Human Factors in Computing Systems}{May 8--13, 2021}{Yokohama, Japan}
\acmBooktitle{CHI Conference on Human Factors in Computing Systems (CHI '21), May 8--13, 2021, Yokohama, Japan}
\acmPrice{15.00}
\acmDOI{10.1145/3411764.3445718}
\acmISBN{978-1-4503-8096-6/21/05}

\usepackage{graphicx}
\usepackage{caption,subcaption}
\usepackage[normalem]{ulem}

\usepackage{subcaption}

\usepackage[inline]{enumitem}

\usepackage{dblfloatfix} 

\usepackage{amsmath}
\usepackage{MnSymbol}
\usepackage{wasysym}

\usepackage{xcolor}

\usepackage{xargs}

\newcommand{\bfemph}[1]{\textit{#1}}
\renewcommand{\emph}[1]{\bfemph{#1}}

\begin{document}

\renewcommand{\labelitemi}{\tiny$\blacksquare$}
\renewcommand\labelitemii{\tiny$\square$}

\newcommand*\quotes[1]{``#1''}

\newcommand{\app}{iCoRA}

\newcommand{\mode}[1]{{\fontfamily{zi4}\selectfont#1}}

\newcommand{\college}{ESPOL}

\newcommand{\collegeurl}{\url{http://www.espol.edu.ec}}

\newcommandx{\note}[2][1=]{\todo[linecolor=red,backgroundcolor=red!25,bordercolor=red,#1]{#2}}
\newcommand{\ggm}[1]{\textcolor{purple}{#1}}

\newcommand{\appDefinition}{\textbf{i}nteractive \textbf{Co}urse Selection and \textbf{R}ecommendation \textbf{A}ssistant}

\newcommand{\commentl}[1]{\textcolor{teal}{#1}}
\newcommand{\remarkl}[1]{\textcolor{green}{Luis: #1}}

\newcommand{\todonote}[1]{\textcolor{red}{\textbf{#1}}}

\title[Showing Academic Performance Predictions during Term Planning]{Showing Academic Performance Predictions during Term Planning: Effects on Students' Decisions, Behaviors, and Preferences}




\author{Gonzalo Gabriel M{\'e}ndez}
\affiliation{%
  \institution{\hbox{Escuela Superior Polit{\'e}cnica del Litoral}}
  \city{Guayaquil}
  \country{Ecuador}}
\email{gmendez@espol.edu.ec}

\author{Luis Gal{\'a}rraga}
\affiliation{%
  \institution{INRIA}
  \city{Rennes}
  \country{France}}
\email{luis.galarraga@inria.fr}

\author{Katherine Chiluiza}
\affiliation{%
  \institution{\hbox{Escuela Superior Polit{\'e}cnica del Litoral}}
  \city{Guayaquil}
  \country{Ecuador}}
\email{kchilui@espol.edu.ec}



\renewcommand{\shortauthors}{M{\'e}ndez, Gal{\'a}rraga \& Chiluiza}

\newcommand{\ommitl}[1]{\textcolor{red}{\sout{#1}}}

\newcommand{\rev}[1]{\textcolor{blue}{#1}}

\begin{abstract}
  Course selection is a crucial activity for students as it directly impacts their workload and performance. It is also time-consuming, prone to subjectivity, and often carried out based on incomplete information. This task can, nevertheless, be assisted with computational tools, for instance, by predicting performance based on historical data. We investigate the effects of showing grade predictions to students through an interactive visualization tool. A qualitative study suggests that in the presence of predictions, students may focus too much on maximizing their performance, to the detriment of other factors such as the workload. A follow-up quantitative study explored whether these effects are mitigated by changing how predictions are conveyed. Our observations suggest the presence of a framing effect that induces students to put more effort into course selection when faced with more specific predictions. We discuss these and other findings and outline considerations for designing better data-driven course selection tools.

\end{abstract}




\maketitle

\section{Introduction}

Academic advising is a crucial aspect in the mission of any Higher Education Institution (HEI). One of its central components is course recommendation. This is of paramount importance to students as proper course selection has a direct impact on their academic workload and overall performance \cite{Bar2005QuestFK}.

Course recommendation is usually performed by a designated advisor who assists students in selecting the most appropriate courses for their upcoming term. This advice is based on the advisor's knowledge of the academic program and its history, as well as her ability to craft personalized recommendations from that information. The latter factor makes academic advising particularly challenging: Since each student's history and profile is unique, advisors are repeatedly challenged with previously unseen scenarios that require a thorough analysis. 
The challenge becomes tougher as advising must often occur within a short period of time~\cite{grupe2002internet}, making course recommendations prone to errors and susceptible to subjective views.
For example, an individual's learning experience may likely influence her perception of the difficulty of a given course. As students lack a global view of the study program, they also tend to make decisions based on the \emph{vox populi}.



For all the reasons mentioned above, some efforts aim at assisting academic advising with data-based visualization tools (e.g., ~\cite{guttierez2018LADA, Charleer:2018:LearningAnalyticsDashboards}). The goal of such tools is not to replace the human advisor but to empower both students and advisors with complementary actionable advice based on a \emph{more objective} view of the students' enrollment alternatives. That view can be based on official information about the study program (e.g., the courses' number of credits, their expected workload), the historical difficulty of the courses, and the student's historical performance.

A key aspect when designing such data-based tools is how to characterize the performance of students. This is due to the fact that the chosen metric may steer the students' attention to specific aspects of their professional instruction. 
The GPA, for example, is often seen as a key factor for success in the labor-market and there is a cultural tendency to frame college students based on it~\cite{ting2012understanding}. For this reason, it is not uncommon for students to try to maximize their GPA regardless of their actual development of knowledge, skills, or understanding~\cite{fleur2020learning}. Hence, the GPA has limitations in reflecting a student's academic performance. From a pedagogical perspective, performance metrics that seek to assess and develop the \quotes{21st century skills}~\cite{rotherham201021st} (critical thinking, collaboration, creativity, long-life learning, etc.), are more desirable. However, metrics of this kind are seldom collected by HEIs in a systematic fashion. Ultimately, the use of a specific performance metric in a data-based tool must observe any availability constraint. That is, it is dependent on the metrics readily available at the HEIs.

Regardless of the selected performance metric, visualization tools are promising for student-oriented course recommendation because such tools can efficiently convey the multiple facets of a study program. In this line of thought, we present the findings of two studies carried out with iCoRA (\appDefinition), a tool that supports students in deciding their upcoming term's enrollment, prior to their planning advising meeting. \app{} is part of an initial effort to improve the course recommendation process at the Escuela Superior Polit\'{e}cnica del Litoral (\college), a Latin-American university. The tool's recommendations are based on course grade predictions, which are computed by integrating the available information at \college{}, namely the student's grades and data about the courses such as workload, number of credits, pre-requisites, and historical performance. \app~also provides explanations for its predictions.



The two studies conducted with \app~required students to compose and decide on a set of courses for their upcoming term. We first conducted a qualitative study that investigated the effects of showing performance predictions on the students' decisions. Here, the grades predicted by \app~were presented through a range-based visual representation. We found that in the presence of these predictions, students focused mainly on maximizing the predicted grades, paying less attention to other important factors that may play a role in their term outcome (e.g., the workload).
This aligns with the results of previous research on the unintended consequences of exposing students to historical performance data based on the GPA (e.g.,~\cite{Chaturapruek2018, Bar2005QuestFK, Aguilar2020associations, sample2018implications}). We argue that this type of overreliance effect constitutes an important limitation 
of making GPA-based predictions.

In a follow-up quantitative study, we then investigated whether the effects observed in our qualitative evaluation could be mitigated \textbf{through design}, by changing the visual representation of the predictions. To this end, we modified \app~to convey its predictions through eight different visual representations that span a \emph{specific} to \emph{vague} spectrum. This study focused on characterizing not only the students' decisions, but also their decision process and preferences. We found that some visual representations had significant effects on the students' chosen workload and the time they interacted with the tool's explanations for the predicted grades.


This paper contributes empirical evidence on the impact that grade- and GPA-based predictions have on the behavior of students, as well as the role played by the visual representations of those predictions. We discuss our findings in the context of \app{} and \college{}, not without arguing the context-related 
limitations of our design choices, and the identified effects of showing grade predictions to students. The paper also contributes a discussion on the potential ethical concerns that may arise from providing students with GPA-based predictive tools to support their enrollment decisions. Based on all of this, we devise several considerations and potential principles for the design of new effective data-driven tools for course selection and recommendation.

\section{Related Work}
\label{sec:relatedwork}




Course selection and academic performance prediction are often discussed within the realm of Learning Analytics (LA)~\cite{LongSieme2014hy}. In this section, we first review existing student-oriented visualization tools in the LA literature. Since \app's recommendations are based on grade predictions, we then survey studies on the effects that exposing students to GPA and historical performance information has on their enrollment decisions and behavior. We conclude this section with the state of the art in visualization design choices and how these affect viewers' interpretation of visual representations. We build upon knowledge from these areas to inform the design of \app~and the studies we present in this paper.


\subsection{Visual Learning Analytics and Tools for Academic Advising}
Viera et al.~\cite{Viera2018_VLA} use the term Visual Learning Analytics (VLA) to refer to LA and Educational Data Mining (EDM) techniques that are facilitated through interactive visual interfaces. Defined as ``\textit{the use of computational tools and methods for understanding educational phenomena through interactive visualization techniques}''~\cite[p.~120]{Viera2018_VLA}, this research area lies at the intersection of LA, EDM, and Information Visualization (InfoVis).

In the area of academic advising, LISSA~\cite{Charleer:2018:LearningAnalyticsDashboards} and LADA~\cite{guttierez2018LADA} are notable examples of VLA tools. LISSA uses historical data to predict the probability of graduation of students within the career's expected time. This is used by advisors to plan enrollment of first-year students who have previously failed courses. Using clustering techniques, LADA predicts the probability that a student fails a course.

Both LISSA and LADA target teachers and advisors as their final users. Student-oriented advising tools are less common. One relevant example in this category is KMCD~\cite{4722942}, a self-advising system that shows courses for enrollment based on a given curriculum design. CARTA \cite{4722942} is another course planning tool that provides students with course descriptive information, evaluations of instructors, and grade distributions. \app~shares with KMCD and CARTA the goal of making information on historical data of courses available to students. However, in line with known guidelines for student-oriented VLA tools (e.g.,~\cite{8010828, 10.1145/2883851.2883904,10.1007/978-3-642-40814-4_51,10.1145/2883851.2883888,atapattu2016topic,4722942}), \app~resorts to visualization techniques to also provide performance predictions through visual representations.

\subsection{Exposing Students to Historical Performance Information}

Several research efforts have investigated the impact of disclosing information about performance of previous cohorts on students. According to Ognjanovic et al.~\cite{Ognjanovic2016}, the knowledge of historical GPAs is a key factor to explain the courses students opt for. It has been found that when students have access to the performance outcomes of previous courses, they tend to choose leniently graded courses~\cite{Bar2005QuestFK} or make shortsighted choices regarding their careers~\cite{Smith1995UnintendedConsequences}. In a more focalized context, Lim et al.~\cite{Lim2019} found more recently that even Learning Analytics Dashboards (LADs) may have a negative impact on students because of the social anxiety they experience when their peers performance is compared to theirs. These and other unintended consequences~\cite{Chaturapruek2018} often prevent HEIs from making performance and GPA information publicly available. 

On the other hand, a parallel line of research found that when students are indirectly exposed to academic performance visualizations through their advisors or counselors during one-to-one meetings, they show---over a relatively short period of time---positive changes in motivation and self-regulated strategies for learning~\cite{Aguilar2020associations}. Along the same lines, Main \& Ost~\cite{Main2014ImpactOfLetterGrades} identified that there was no evidence of the effect of letter grades on the students' enrollment decisions. They also found a positive effect on the students' efforts within courses. 

The body of work referred above suggests that there is still the need to study the impact of LADs and data-based tools that expose students to historical performance information. We take steps in this direction with a special focus on visualization, by also investigating the role that different visual representations play when presenting performance predictions to students.

\subsection{Frames and Visual Representations}

A framing effect arises when people make different choices based on how a given problem---or set of options---is presented. This type of cognitive bias has been widely studied in opinion formation (e.g.,~\cite{chong2007theory, nelson1999issue, druckman2001implications}) and decision making processes (e.g.,~\cite{mcelroy2003framing, levin1988information, levin1998all}). Framing effects have also been observed when visual representations are used as communicative structures of a message. Cheema et al.~\cite{Cheema2011EffectGoalVisualizationOnGoalPursuit} found that visual representations for goal progress (e.g., progress bars) enhance motivation as people approach their goal. Low-level visual features such as spacing, position, and order have also been found to impact the responses elicited by survey questions as well as the response process~\cite{Tourangeau2004SpacingPositionOrder}. Baumer et al. explored how framing effects can be mitigated in text visualizations of political issues~\cite{baumer2017simple, Baumer2018}. Other explorations in the context of human rights narratives have investigated the effects that anthropomorphizing standard charts has on the empathy and prosocial behavior of the viewers~\cite{Boy2017PeopleBehindData}.

In a broad sense, rhetorical techniques---the choices made at the data, visual representation, annotation, and interactivity levels---steer our thinking of the topics presented by a visualization. In consequence, those techniques affect end-user interpretation~\cite{Hullman2011}. The way different design choices prompt viewers to interpret visualizations from different perspectives has been investigated at different levels: from the effectiveness of low-level visual mappings~\cite{Cleveland1984, Cleveland1987, WareBook}, to the impact of more high-level concepts and visualization elements such as titles~\cite{Kong2018FramesAndSlants, Kong2019TrustAndRecall} and visual embellishments~\cite{Bateman}. The impact of the latter group has been studied in different contexts: visualization recognition, recall, comprehension and interpretation, memorability, perception of bias, and change of attitude. All these are important processes that viewers experience when exposed to visual representations of data.

Inspired by the body of knowledge summarized above, we are interested in investigating the effect that different visual representations of academic performance predictions have on the students' decisions when planning their upcoming terms. In this work, we define a continuum of prediction representations---ranging from \textit{specific} to \textit{vague}---and study how the students' decisions, decision processes, and preferences are shaped by these representations.











\section{Motivating Context and Research Questions}

\college\footnote{\collegeurl} is an engineering-oriented Ecuadorian university with over 10,000 students and 32 undergraduate programs. The advisors of its academic advising system are lecturers chosen by workload availability who are assigned up to 40 students (25 on average). Advising sessions take place twice every term over a two-week period: right before the term begins (for course selection and recommendation) and after the midterm exams (to monitor the students' performance). Each advising session is supposed to last no longer than 15 minutes.

In-house observations and interviews revealed that it is common for students to arrive unprepared or undecided to their term planning advising appointments. This makes advising sessions longer, which is particularly problematic when the students have other issues that also need to be addressed during the meeting. Besides, this lack of preparation may induce the students to select their courses on the spot, likely on the basis of unofficial, incomplete, and potentially non-accurate information. For instance, in-house inquiries about the activities students perform to decide on their courses, reveal that 83\% ask other fellow students not only about the difficulty of the courses, but also about the reputation of lecturers. These inquiries also indicate that students deem their fellow students' advice as important as their advisor's. 


\begin{figure*}[b]
     \centering
  \includegraphics[width=\linewidth]{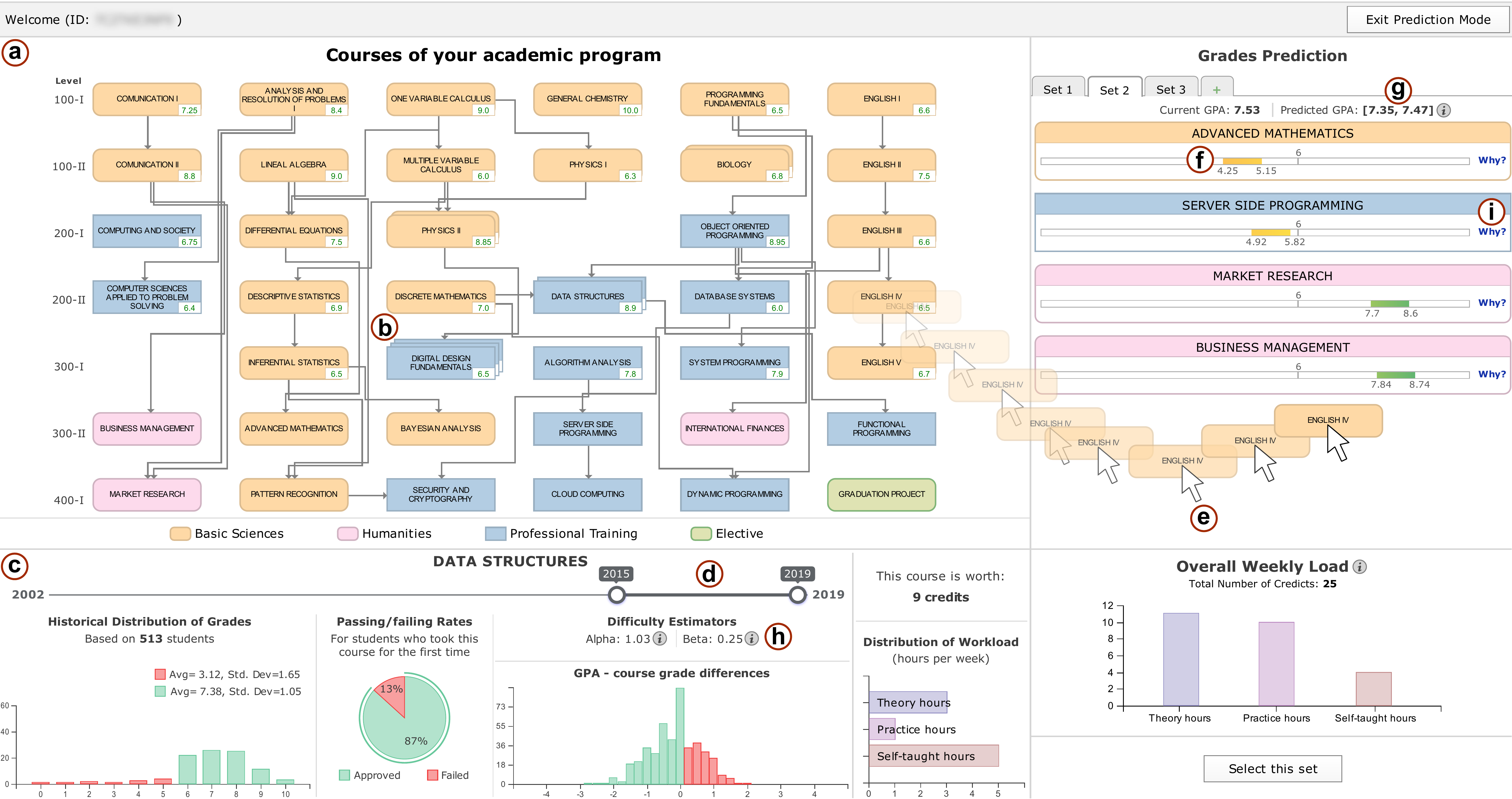}
  \caption{\app's student module. The main view shows the courses of the student's academic program. Clicking an element of this view reveals the course' history and general information. Under the prediction mode, courses can be dragged onto the prediction panel. In response to these interactions, \app~predicts the student's performance in each of the selected courses.}
  \Description{The interface of \app~is shown here. This includes a representation of a CS academic program in which courses are organized on a grid structure. Connections between courses represent pre- and co-requisites. The image also contains several visualizations showing the historical information of the course, such as the historical distribution of grades and the passing and failing rates. The interface also includes a panel where courses can be dragged to trigger the tool's prediction models.}
  \label{fig:fullInterface}
\end{figure*}

At \college, students pass a course with a minimum grade of 6.00 (out of 10) and are ranked in terms of their GPA, which is reflected in their official academic record and transcripts. Although some instructors may conduct class activities using alternative performance metrics (e.g., development of learning outcomes, levels of engagement), \college's current grading policy enforces all evaluations to be captured via the students' grades and, consequently, their GPA. This information is also commonly requested by recruiters of the local market in job applications. For these reasons, the students at \college{} deem GPA performance highly important, so much so that they carefully consider any potential impact on their GPA when making enrollment decisions.\\

The aforementioned observations suggest that a tool that supports the data analysis aspect of course selection could help students not only in preparing for their term planning appointments, but also in making more informed decisions. Ultimately, this could alleviate the advisor's workload and provide students with a more objective view of their study program. We highlight that, rather than replacing the advisor, a tool of this type has the potential to make the student-advisor dialogue more effective and efficient. However, before such a tool could be deployed in a real-world setting, we would need to understand:\\

\textbf{RQ1:} What are the effects of showing performance predictions to students during term planning?

\textbf{RQ2:} How do these effects vary when we change the visual representations used to convey the predictions?\\

We investigate these questions through the lens of \app~\cite{Castells2020}, an interactive visualization tool that provides students with historical data on their academic program. The studies conducted with \app~focus on the Computer Science (CS) program of \college, that is composed of 41~courses (104~credits). 37~of these courses (96~credits) are compulsory while the remaining 4~(8~credits) are elective. This curriculum design makes the enrollment less flexible than most universities in Europe and North America, where students can often \textit{mix and match} a wider variety of courses based on their interests and tastes.

Given the extensive use of the GPA at \college{} and its importance for the students' career prospects, \app's current implementation issues course recommendations based on the students' past grades. That said, we do acknowledge the limited capacity of the GPA to fully describe a student's learning, capabilities, and skills. A myriad of factors beyond course grades have shown to influence students’ performance (e.g., demographic and socio-economic background~\cite{Green2011StudentDemographic}; high school history~\cite{klomegah2007predictors}; social ties with classmates~\cite{Gasevic2013}; personality and psychological aspects such as self-efficacy~\cite{Bandura1994}, motivation~\cite{Ryan2000WhenRewardsCompeteWithNature, Ryan2000SelfDetermination}, and approaches to learning and preferences for teaching/courses~\cite{Chamorro2008}). Therefore, this investigation should be regarded as evidence of the effects of exposing students to performance predictions \textit{in general}. Our goal is to provide a reference for the design of tools based on other performance metrics, by showing how certain design choices may shape the students' behavior (see also section~\ref{subsec:ethical_considerations} in the Discussion).

\section{\app}
\label{sec:system}


\app~\cite{Castells2020} is a tool that assists students in planning their upcoming term in preparation to their advising appointments. It supports the composition of arbitrary sets of courses available for enrollment. Based on past observations, it provides performance predictions and information on the term's resulting workload and difficulty.\\

Although \app~is not the main contribution of this paper, this section describes the tool in detail as its components are relevant for the studies later described.







\subsection{Students' Academic Program and History} 
The \emph{program view} shows the student's academic program as a grid of courses with links indicating pre- and co-requisites (Figure~\ref{fig:fullInterface}a). Courses are organized into four categories (basic science, professional training, humanities, and elective) and are color-coded accordingly. This view shows each course with the grade obtained by the student; the grades are shown in green for passed courses, and in red for failed ones.
Courses that have been repeated are depicted as groups of stacked rectangles, each representing an enrollment instance (e.g., Figure~\ref{fig:fullInterface}b).\\

Clicking on a course of the program view displays the course's general and historical information  (Figure~\ref{fig:fullInterface}c): number of credits, weekly workload, difficulty estimators (course grading standard $\alpha$ and grading stringency $\beta$---as defined in~\cite{Caulkins:1996:AdjustingGPA}), distribution of grades, and historical performance. This data can be filtered by time through an interactive range slider (Figure~\ref{fig:fullInterface}d). This supports the exploration of the course's evolution over time and provides insights about the performance of students who have recently enrolled in a given course. This is relevant to support students in making decisions in light of recent data. 








\subsection{Course Sets and Performance Predictions}



Under the \emph{prediction mode}, available courses from the program view can be dragged onto the \emph{grades prediction} panel (Figure~\ref{fig:fullInterface}e) to compose one or more sets of courses. These interactions trigger the execution of \app's performance prediction models and update the panel's content.

The prediction models for each subject are based on gradient boosting trees (GBT) trained on historical data that comprise term workload, previous grades, failing history, and aggregated course difficulty. In the version of \app~shown in Figure~\ref{fig:fullInterface}, the performance prediction of each course is depicted as a range---computed via quantile regression on GBT---on a horizontal scale between 0 and 10, in compliance with \college's grading system (Figure~\ref{fig:fullInterface}f).
The range is shown through a red-yellow-green divergent color scale with a zero value of 6.00---the minimum passing grade.

On adding courses to---and removing them from---the prediction panel, \app~estimates the student's GPA that would result if the predictions shown became true. The GPA is estimated by considering the lower and upper bounds of the ranges predicted and is presented on the interface also as a range (Figure~\ref{fig:fullInterface}g).

\subsection{Explanations}

\begin{figure}[t]
\centering
\includegraphics[width=0.47\textwidth]{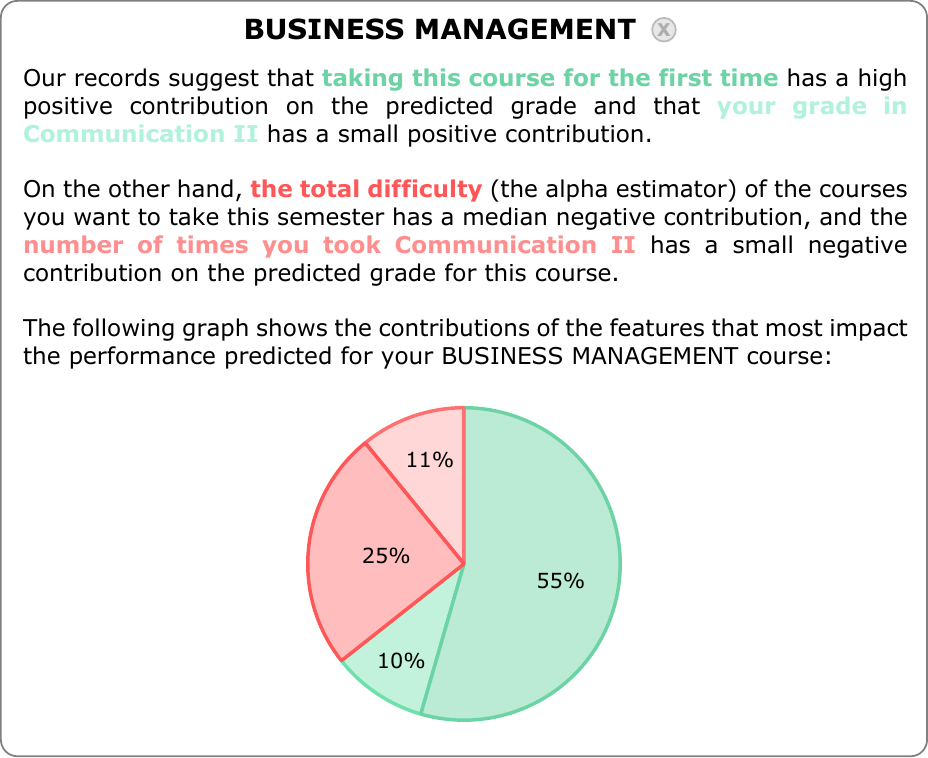}
\caption{Explanation of the grade predicted for a \textit{Business Management} course. Besides the pie chart, the version of \app~used in our first study included a written summary of the impact of the model's input features.}
\label{fig:explanation:study1}
\Description{A pie chart shows the relative distribution of the input features that impact the performance predicted for a Business Management course. Above the pie chart, there is a textual description of the features' impact. Both the in the textual description and in the pie chart, features with positive influence appear in green while those with a negative impact appear in red.}
\end{figure}

\app~provides explanations of some of the features used by its prediction models.  
These explanations combine text, very simple visualizations, and math formulas. Examples include tooltips describing the difficulty estimators for courses (Figure~\ref{fig:fullInterface}h). The performance predicted for each course is also explained. The \textit{Why?} button to the right of each prediction (Figure~\ref{fig:fullInterface}i) explains the relative contribution of the model's input features to its output (see Figure~\ref{fig:explanation:study1}). This contribution is calculated with SHAP~\cite{NIPS2017_7062}, an explanation method based on linear feature attribution.

\app~offers these explanations so that the students can capitalize on the factors that could positively influence their performance. Perhaps more importantly, these explanations seek to encourage students to mitigate potential negative impact. For example, a way to reduce the risk of getting bad grades could be to decrease the overall grading stringency (total $\beta$) of the courses. This could be done by enrolling in fewer courses or by taking easier ones.\\

Having introduced the functionalities of \app, we are ready to elaborate on the user studies we conducted with the tool to answer our research questions. These studies investigate the impressions of the students with regard to \app's functionalities, and in particular, shed light on the impact of performance predictions on the decisions, behaviors, and preferences of students in the context of course selection.

\section{Study 1 --- Showing performance predictions}
\label{sec:study}

We ran a qualitative study that investigated \textbf{the effects that showing performance predictions has on students when they plan their upcoming term (RQ1)}. Our original experimental design was based on a controlled lab study. However, the sanitary crisis around the COVID-19 pandemic forced us to convert our protocol into a remote format. We thus used video conferencing software to test and interview participants remotely.


\subsection{Participants and Procedure}

We recruited 12~participants from \college' computer science (CS) undergraduate program (4~female; 8~male; 21--30 years old; median age 24). Students were at different stages of their degree: first (n=2), second (n=6), and third (n=4) year. All had attended at least two academic advising meetings.

In each individual study session, participants were asked to put themselves in the place of the fictional student whose academic history and set of available courses were shown in \app. Participants had to select a set of courses for their upcoming semester, and were allowed to use \app~to work on this task for as long as they wanted. We did not specify restrictions regarding the number of courses they were allowed to take.


Participants worked with a modified version of \college's CS program, where the last two semesters were replaced with courses from other CS curricula. This was enforced by ethics regulations in order to avoid influencing the students' attitude towards actual courses they had not taken yet. 
The introduced courses were chosen so that they seemed plausible,
that is, they had names that participants could understand and relate to (e.g., \textit{Dynamic Programming}).

Participants had to choose among a set of nine introduced courses distributed across basic sciences, humanities, and professional training. Each category had three courses of low, average, and high difficulty.

For the sake of the study, \app~was fed with synthetic data. Grades and aggregated difficulty estimators were randomly drawn from different normal distributions skewed according to the courses' difficulty. A student's failing history (number of times a course was taken) was generated using a power-law distribution. The models that predicted the performance of the surrogate courses consisted of handcrafted linear functions that allowed us to control the contribution of each feature to the predicted performance intervals.

\subsection{Data Collection and Analysis}

We used online questionnaires to collect participants' consent, demographic information, and data on the strategies they usually follow when choosing their courses. We recorded all the sets of courses composed by the participants. We also captured their interactions with \app~through a video conferencing tool. In a post-task questionnaire, participants rated a series of propositions about \app. The interviews were recorded, fully transcribed, and qualitatively coded following a thematic analysis approach~\cite{boyatzis1998}. Our initial coding was done by two researchers independently and focused on the students' general perception and rationale. Higher-level topics emerged in subsequent meetings in which the coding scheme was revised iteratively by the two researchers until a unified coding scheme was reached.





\subsection{Results}
\label{sec:findings}

\begin{figure}[t]
\centering
\includegraphics[width=0.47\textwidth]{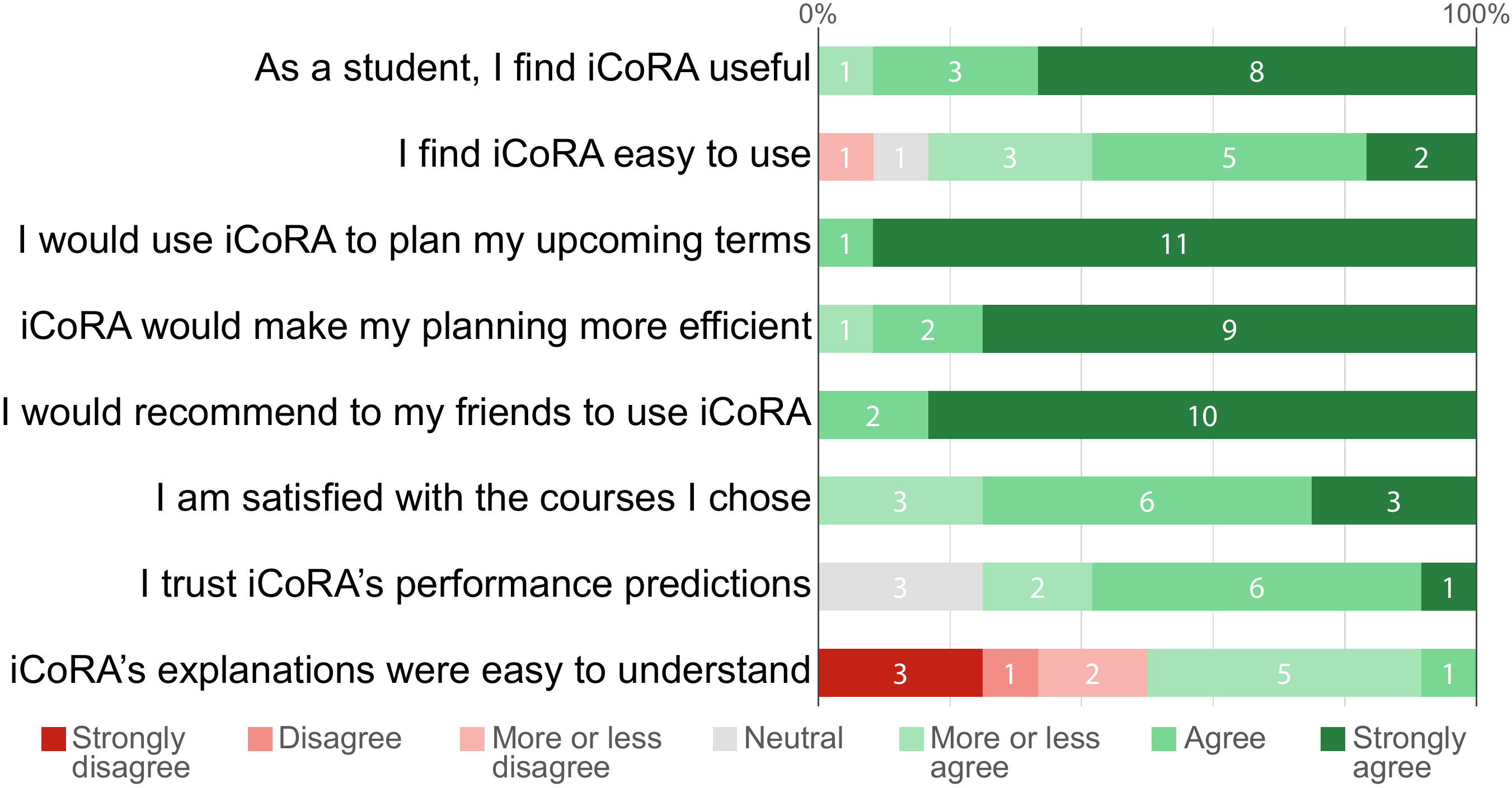}
\caption{Participants' ratings on several propositions about their experience with \app. Each proposition was rated using a 7-point Likert scale.}
\label{fig:likert_scale_icora}
\Description{Participants' ratings on several propositions about their experience with \app. The results for each proposition appear as a horizontal bar with sections whose width depict the proportion of participants for a given rating. Low values appear in red while higher ones appear in green.}
\end{figure}

Our analysis of the questionnaires and interviews revealed a general enthusiasm for \app. Participants particularly appreciated the access to their courses' historical information and highlighted the usefulness of this feature to get an overall impression of a course's reputation---instead of having to ask other fellow students about this. The prediction feature was highly appreciated---both at the course and the GPA levels. Figure~\ref{fig:likert_scale_icora} shows a summary of the participants' ratings of several aspects of \app. These results are displayed in a 7-point Likert scale.

In the subsections that follow, we present the most important findings of this study. The quotes included below have been translated from Spanish.





\subsubsection{Students' Decisions}
\label{subsubsec:influence_decisions}
Before the course selection task, we asked participants about the strategies they usually follow to decide on their courses. Their answers included aspects such as aiming at a specific term workload (\quotes{\textit{I choose between five and six courses per semester}}~[P01]; \quotes{\textit{I always choose four courses}}~[P11]), balancing the difficulty of their courses (\quotes{\textit{I have to choose this course [...] with easier courses}}~[P03]), and following the sequence in which courses appear in their academic program (\quotes{\textit{I usually don't choose courses from distant levels}}~[P01]).\\

After composing and choosing a course set with \app, during the interviews, we asked students on the rationale behind their decisions. All participants, with no exception, considered the predicted performance of the courses as the most important factor to select their courses: \quotes{\textit{I noticed the grades were better in my second set of courses. So, I chose that.}}~[P03]; \quotes{\textit{[\app] showed me the minimum grade I was going to get and that's important because it affects my GPA for the next semester.}}~[P11]. In five occasions, performance was also mentioned in regards to the predicted GPA: \quotes{\textit{It showed me how my GPA was going to improve by the end of this semester}}~[P07].\\

These statements suggest that \app's predictions heavily influenced participants' approach to course selection. The tool seemed to have turned the participants' attention to the predicted grades, away from other aspects that students traditionally consider when deciding on their courses. We found that, in the presence of performance predictions, students perceive course selection as a grade maximization problem. The data supports this hypothesis: The set of courses selected by the students are, on average, at the 96-th percentile in terms of GPA's predicted upper bound when we consider all the course sets they ever composed. When we look at the GPA's lower bound, and the maximal individual grades, the sets lie at the 77-th and 87-th percentiles respectively.\\


Our video analysis also indicates that when selecting courses, students often disregarded factors such as the workload they would face or the difficulty of the chosen courses.

\begin{figure*}[b!]
  \centering
  \includegraphics[width=\linewidth]{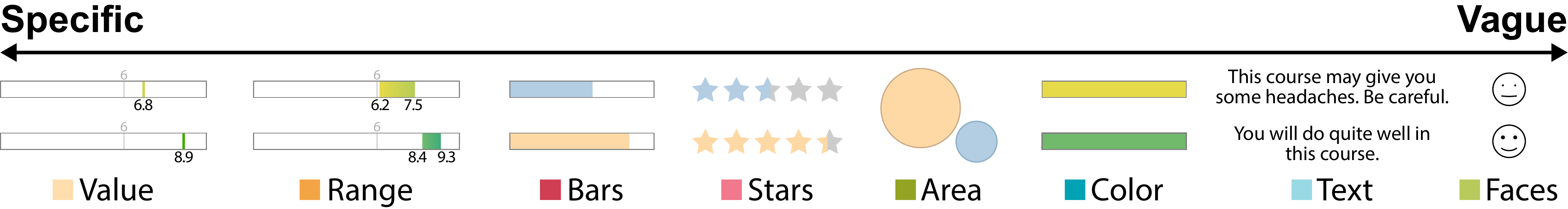}
  \caption{\textit{Specific} to \textit{vague} spectrum of visual representations for performance predictions. See Section~\ref{sec:spectrum} for a detailed description of each representation.}
  \Description{A horizontal graphic shows eight different visualizations lying on a spectrum from specific to vague. Each visualization shows a predicted grade in a different way.}
  \label{figure:spectrum}
\end{figure*}


\subsubsection{Participants' Interest in Explanations}

Explainability is crucial to produce predictions that humans can understand and trust. \app~takes steps in this direction by providing explanations for the course difficulty indicators $\alpha$ and $\beta$~\cite{Caulkins:1996:AdjustingGPA}. In the same vein, the \textit{Why?} button explains the impact of the model's input features on the prediction outcome. This functionality aims at opening the \textit{black boxes} used by the tool. However, the participants' ratings on the explanations suggest that these might have not been very effective. Some students commented on this explicitly: \quotes{\textit{The explanations could be less formal.}}~[P05]; \quotes{\textit{There are too many words, they could be replaced with icons, or perhaps be more concise.}}~[P06]; \quotes{\textit{Show them with other words, they were hard to understand.}~[P11].}\\




The comments of above suggest the version of \app~used in this study has room for improvement regarding how it explains the different pieces of information---which seemed not obvious to the students. 
However, our video analysis also revealed that overall participants interacted very little with the explanations. Regarding the difficulty estimators of the courses, only three participants opened the explanation for $\alpha$ (mean time 8~seconds) and just one checked the explanation for $\beta$ (during 25~seconds). The explanations for the performance predictions provided through the \textit{Why?} button sparked more interest: ten participants opened them at some point, leading to a global average of 1.5~minutes (for the total time). However, the participants' interest in these explanations decreased significantly after their first interaction with them. Only eight participants requested these explanations a second time and the average time they spent on it went from 53 seconds for the first time to just 12 for the second. Further interactions with the \textit{Why?} button were very rare, and always shorter.

\section{Study 2 --- Alternative Visual Representations of Performance Predictions}
\label{sec:study2}


Motivated by the observations presented above, we designed a second study to better understand whether and, if so, how students are influenced by the way \app's performance predictions are conveyed. The driving research question for this study was \textbf{whether different visual representations for performance predictions can affect the decisions and behaviors of the students when they select their courses (RQ2)}. We wanted to see, for example, if text-based performance predictions would make students less eager to maximize their grades. In this study, we investigate these effects not only on the students' final decisions, but also on their decision process and preferences.

We followed a protocol similar to the one of Study 1 but this time, students had to choose courses from several versions of their academic program, each with a different set of available courses. Besides, the performance predictions were displayed using different visual representations. Before describing our experimental protocol in detail, we first explain the alternative visual representations we used to answer RQ2.

\subsection{A Spectrum of Performance Prediction Representations}

\label{sec:spectrum}

For this study, we designed eight different ways to show performance predictions and integrated them into \app. These representations span along a spectrum from \textit{specific} to \textit{vague} (Figure~\ref{figure:spectrum}). This spectrum is inspired in work by Walny et al.~\cite{walny2015Sketching} that describes a continuum of visual representations from countable (numeric) to pictorial (abstract), found by observing how people sketch representations of data. A set of similar representations was found by M\'{e}ndez et al.~\cite{Mendez2017} after comparing the visualization construction process of iVoLVER~\cite{Mendez2016} and Tableau Desktop. Based on these continua, we consider a performance prediction representation to be more \textbf{specific} if it makes the actual grade more directly readable, and more \textbf{vague} if it manipulates the grade to represent it graphically, in a more abstract way. We elaborate on the visual representations that compose our spectrum in the following:

\begin{figure*}[t!]
    \centering
    \begin{subfigure}[t]{0.33\textwidth}
        \centering
        \includegraphics[width=0.9\textwidth]{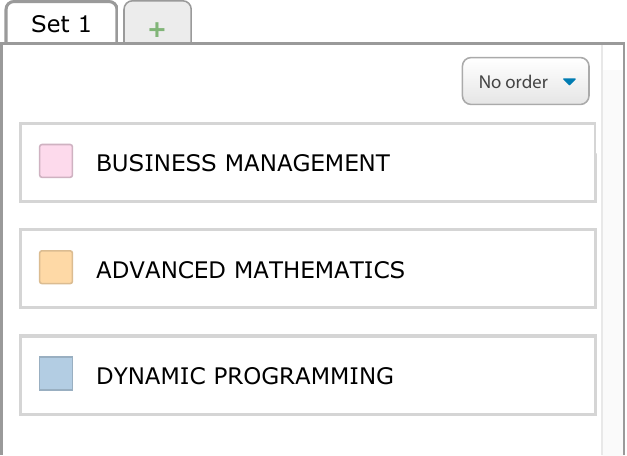}
        \caption{No prediction}
        \label{figure:representation:noPrediction}
    \end{subfigure}%
    ~
    \begin{subfigure}[t]{0.33\textwidth}
        \centering
        \includegraphics[width=0.9\textwidth]{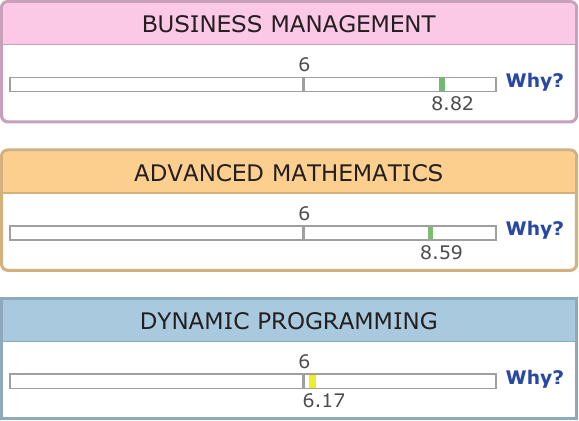}
        \caption{Value}
        \label{figure:representation:value}
    \end{subfigure}%
    ~
    \begin{subfigure}[t]{0.33\textwidth}
        \centering
        \includegraphics[width=0.9\textwidth]{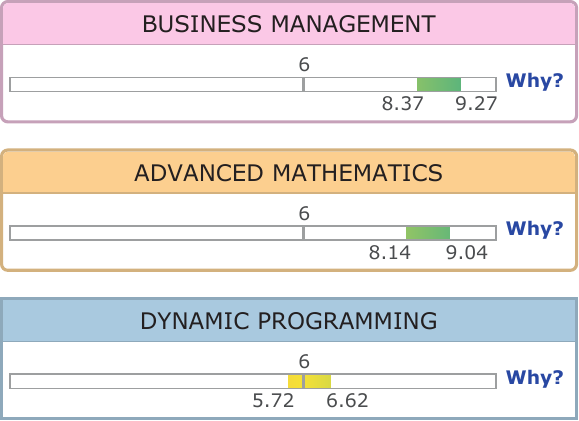}
        \caption{Range}
        \label{figure:representation:range}
    \end{subfigure}%
    \\[0.2in]
    \begin{subfigure}[t]{0.33\textwidth}
        \centering
        \includegraphics[width=0.9\textwidth]{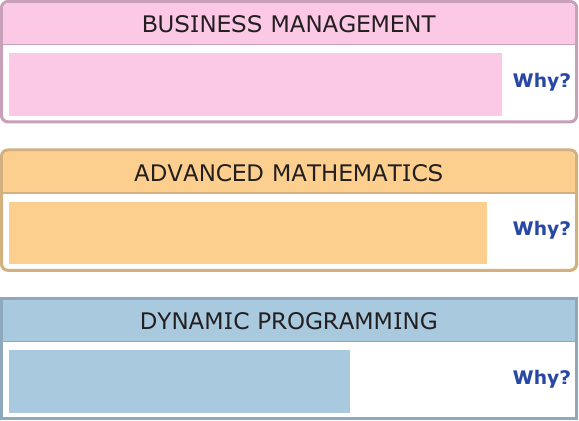}
         \caption{Bars}
        \label{figure:representation:bars}
    \end{subfigure}%
    ~
    \begin{subfigure}[t]{0.33\textwidth}
        \centering
        \includegraphics[width=0.9\textwidth]{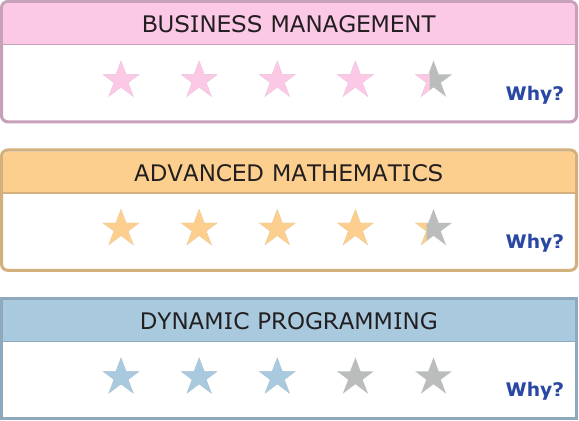}
         \caption{Stars}
        \label{figure:representation:stars}
    \end{subfigure}%
     ~
    \begin{subfigure}[t]{0.33\textwidth}
        \centering
        \includegraphics[width=0.53\textwidth]{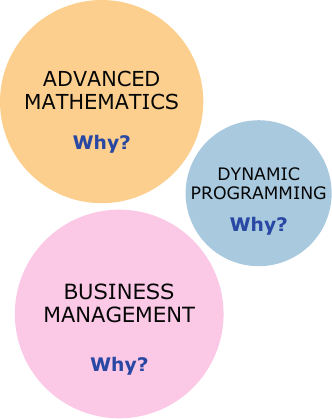}
         \caption{Area}
        \label{figure:representation:area}
    \end{subfigure}%
    \\[0.2in]
    \begin{subfigure}[t]{0.33\textwidth}
        \centering
        \includegraphics[width=0.9\textwidth]{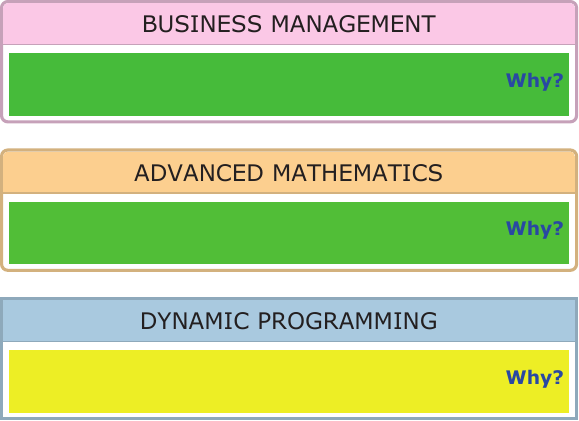}
         \caption{Color}
        \label{figure:representation:color}
    \end{subfigure}%
    ~
    \begin{subfigure}[t]{0.33\textwidth}
        \centering
        \includegraphics[width=0.9\textwidth]{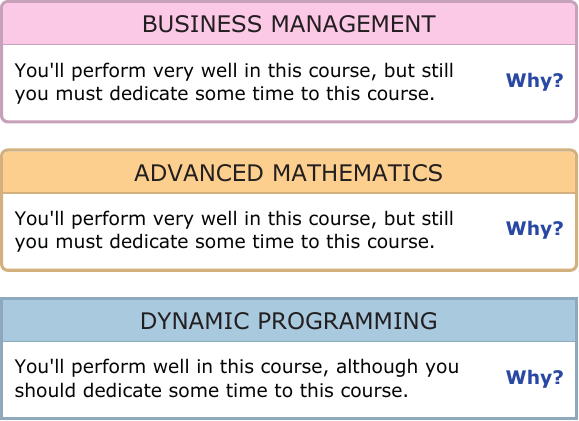}
         \caption{Text}
        \label{figure:representation:text}
    \end{subfigure}%
     ~
    \begin{subfigure}[t]{0.33\textwidth}
        \centering
        \includegraphics[width=0.9\textwidth]{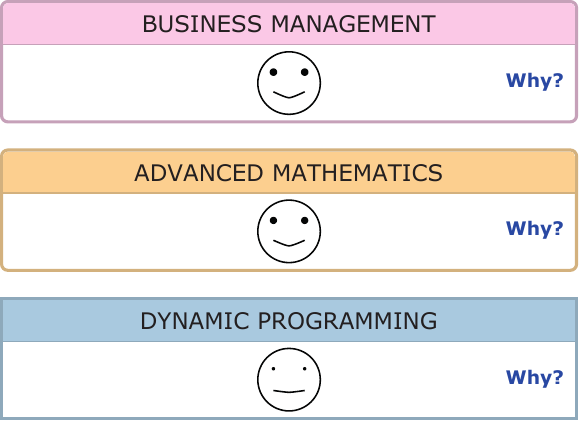}
         \caption{Faces}
        \label{figure:representation:faces}
    \end{subfigure}
    \caption{Performance predictions from the spectrum of Figure~\ref{figure:spectrum} as show by \app~in Study~2. Sub-figure (a) shows how the interface presented a set of selected courses in the control condition (i.e., when no performance prediction was provided).}
    \label{figure:predictionRepresentationsIcora}
    \Description{Several visualizations show how the differen predictions of \app~were shown in Study 2. This image essentially shows how the visual representations of the spectrum previously described looked like in \app.}
\end{figure*}


\begin{itemize}[label={-}, leftmargin=0.1in]

\item {\mode{value}}: Shows the predicted grade with a line mark along a $[0.00-10.00]$ horizontal scale. The line mark is colored according to the red-yellow-green scale of the range representation used in Study 1 (Figure~\ref{figure:representation:value}).

\item {\mode{range}}: This is the range representation used in Study 1. It shows the lower and upper bounds of the interval predicted by our models. It uses the same continuous color scale of Study 1 (Figure~\ref{figure:representation:range}).

\item {\mode{bars}}: Fills a portion of a horizontal bar with color indicating the course type (e.g., humanities). The bars of all the selected courses are aligned, which essentially composes a horizontal bar chart with a common left baseline (Figure~\ref{figure:representation:bars}).

\item {\mode{stars}}: Represents the predicted course grade by filling a set of five stars---similar to those used in rating systems. This representation could be considered as a discrete version of the \verb|bars| one. Color is also used here to depict the course type (Figure~\ref{figure:representation:stars}).

\item {\mode{area}}: Uses circular marks that scale relative to each other to represent the predicted grades of a set of courses. Color is used here to depict the course type (Figure~\ref{figure:representation:area}).

\item {\mode{color}}: Uses full, single-colored bars to encode the grade of each course. The color comes from the red-yellow-green color scale used by the \verb|value| and \verb|range| representations (Figure~\ref{figure:representation:color}).

\item {\mode{text}}: Shows a text qualifying the course's predicted grade. The tone of the message varies from \quotes{\textit{It is very likely that you will fail this subject; you will have to prioritize it over your other courses.}} for grades between $[0.00-2.00)$ to \quotes{\textit{You will do excellent in this subject; your grade may make you look exceptional in relation to other students.}} for grades in the range $[9.00-10.00]$ (Figure~\ref{figure:representation:text}).


\item {\mode{faces}}: Shows a colorless emoji-like face made up of two circular eyes and a curved mouth. As the course grade gets closer to 10, the eyes scale up and the curvature of the mouth increases. This is a very minimalistic version of the Chernoff faces~\cite{Chernoff1973} (Figure~\ref{figure:representation:faces}).


\end{itemize}

These representations aim to cover a wide range of levels of specificity at conveying a predicted grade. The ends of the spectrum represent grades in very different ways: the {\verb|value|} representation is very specific, whereas the {\verb|faces|} representation requires decoding the grade from an abstract representation. We consider the representations to be split equally between the \textit{specific} and \textit{vague} categories: Four of them (\mode{value}, \mode{range}, \mode{bars}, and \mode{stars}) are located at the \textbf{specific} side of the spectrum while the remaining four (\mode{area}, \mode{color}, \mode{text}, and \mode{faces}) lie closer to the \textbf{vague} end. However, we remark that the exact position of each visual representation along the continuum should not be considered definitive. Especially within each category, some representations have similar effectiveness to encode quantitative values~\cite{Cleveland1984, Cleveland1987, MunznerBook}. This is particularly true for the \mode{bars} and \mode{stars} representations of the \textit{specific} category and for the \mode{area} and \mode{color} representations of the \textit{vague} end.

Figure~\ref{figure:predictionRepresentationsIcora} provides examples of predictions using the representations of our spectrum. The example shows the performance predicted for a set of three courses: \textit{Business Management}, \textit{Advanced Mathematics}, and \textit{Dynamic Programming}. Figure~\ref{figure:representation:noPrediction} shows how the tool presented this set of selected courses when no prediction was provided. The remaining ones show the \textit{specific} (Figure~\ref{figure:predictionRepresentationsIcora}b--e) and \textit{vague} representations (Figure~\ref{figure:predictionRepresentationsIcora}f--i).


\subsection{Experimental Design}




For Study 2, we followed a between-group design with respect to the \textbf{prediction representation type} (\textit{specific} and \textit{vague}). Each of these independent variables has four levels: \mode{value}, \mode{range}, \mode{bars}, and \mode{stars} for the \textit{specific} condition; and \mode{area}, \mode{color}, \mode{text}, and \mode{faces} for the \textit{vague} one.

For each of these two conditions, we had two dependent variables: student's decisions and behavior. The students' decisions (i.e., the course set they chose) are operationalized via four dimensions: the \textbf{number of selected courses}, the \textbf{average predicted grade} of those courses, the course set's \textbf{average workload} (expressed in hours per week), and its \textbf{total workload}. On the other hand, the students' behavior during a course selection task was operationalized through the time they interacted with the explanations for the performance predictions. More specifically, we measured the student's behavior by the \textbf{number of times} they invoked the tool's explanations, and the \textbf{total time} these explanations remained open.

We then used the levels of the independent variable \textbf{prediction representation type} to conduct a within-subject analysis for each experimental condition. These analyses compared the measurements of the dependent variables within each level (\textit{specific} and \textit{vague}).

\subsection{Participants}


For this study, we invited students of two \textit{Human-Computer Interaction} and one \textit{Data Structures} courses. Out of the 105 students enrolled in these courses, 91 volunteered to participate (74 male, 17 female; 19--32 years old---median 22). All were enrolled in \college's CS undergraduate program and none had participated in Study 1. All have had prior academic advising and were at different stages of their degree: second or third year (n = 39), fourth (n = 27), and superior years (n = 25).




\subsection{Procedure}

We modified the version of \app~used in our first study to display the sequence of forms, tasks, and questionnaires that participants had to work with. We made this modified version of the tool available online. It had a wizard-like interface design that guided participants through the following sequence of activities:



\paragraph{Introduction to \app}
After providing consent and filling out a questionnaire about their demographics and course selection habits, each participant watched a 12-minute video that explained \app's user interface. The video described how to compose sets of courses, and the tool's performance predictions and explanations. It also elaborated on the tasks participants had to complete. 

\paragraph{First course selection task: No Prediction mode}
In this study participants had to complete five course selection tasks, always starting with a scenario in which \app~did not display any performance prediction (as shown in Figure~\ref{figure:representation:noPrediction}). Similar to the procedure of Study 1, participants were instructed to put themselves in the shoes of the student whose academic history and set of available courses were presented. They had to compose a set of courses to enroll in \textit{their} upcoming term and submit their selection.

We introduced the \textit{no prediction} mode in \app~in order to familiarize the users with the tool before being exposed to performance predictions. Furthermore, this condition provided us with a control scenario that allowed us to contrast the effect of the mere presence of performance predictions on the users, regardless of the chosen visual representation. This course selection task was followed by a questionnaire on the rationale behind the participants' decisions.


\paragraph{Four course selection tasks with prediction}
The \textit{no prediction} course selection task was followed by four others, each of which presented the predicted grades through the visual representations of a single type (\textit{specific} or \textit{vague}). Due to our within-subject study for the representation type, each participant was exposed either to the \textit{specific} visualizations or to the \textit{vague} ones. The association participant-representation type was done randomly, before the execution of the study. We used Latin squares to balance the order in which each participant saw the corresponding representations.

All the course selection tasks were based on the same academic program and the history of the same fictional student. However, the courses available for enrollment differed among tasks. Following the strategy used in Study 1, we introduced courses from other CS curricula at semesters six and seven of the academic program shown to our participants. 9 out of the 11 introduced courses were available for enrollment and were distributed uniformly among the three course categories defined by \college. Each category contained a hard course, one of average difficulty, and an easy one. The introduced courses were unique to a selection task. That is, in every task, participants would see a different set of available courses that had not appeared before and would not appear in subsequent tasks. Under the hood, however, the set of available courses was the same in terms of type, prerequisites, workload, difficulty, historical distribution of grades, and underlying prediction model. Only the names and the position of the courses within the program were different. For example, the course \textit{Micro- \& Nanotechnologies} that appeared in the \mode{value} visual representation had the same features as the \textit{Data Protection} course of the \mode{text} representation. We made this decision to make the students' chosen sets comparable across different course selection tasks.


Each course selection task was followed by the same rationale
questionnaire used after the \textit{no prediction} task.

\paragraph{Closing questionnaire.} The experiment concluded with a final questionnaire asking participants about their
preferences on how \app~presented its predictions.\\

\begin{figure}[t]
\centering
\includegraphics[width=0.47\textwidth]{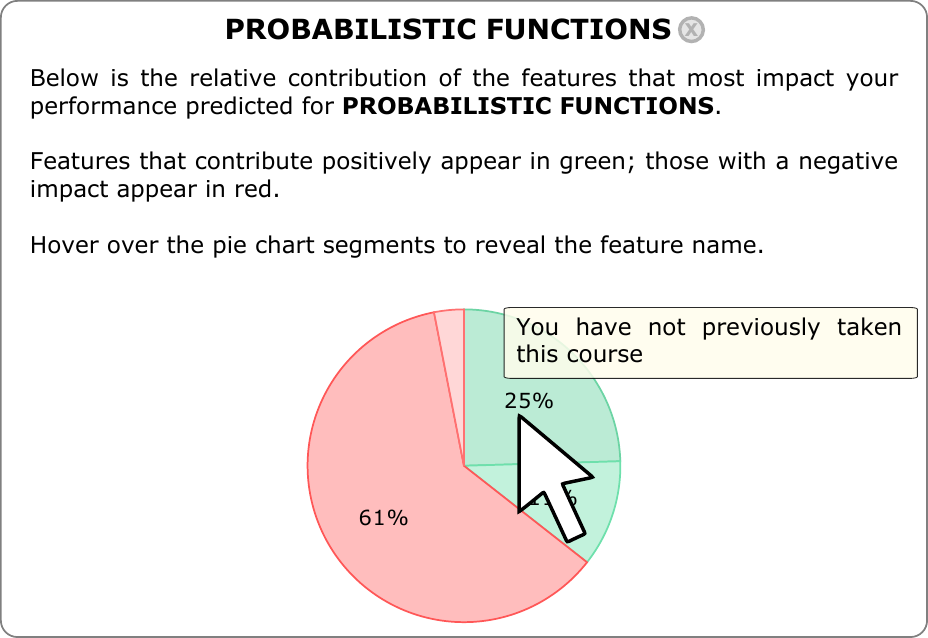}
\caption{Simplified explanation shown by the \textit{Why} button in the version of \app~used in Study 2}
\label{fig:explanation:study2}

\Description{A pie chart shows the relative distribution of the input features that impact the performance predicted for a Probabilistic Functions course. In the pie chart, features with positive influence appear in green while those with a negative impact appear in red. No textual summary is provided above the pie chart.}

\end{figure}

Based on the observations and participants' comments of Study 1, for this study we simplified the explanations shown by the \textit{Why} button. Specifically, we removed the textual summary. Figure~\ref{fig:explanation:study2} depicts how \app's prediction explanations looked like in this study.
We also hid from the interface the section that shows the changes in the student's GPA (Figure~\ref{fig:fullInterface}g) in order to study the influence of the predicted grades in isolation.

\begin{table*}[hb]
\centering

    \begin{subtable}[h]{0.37\textwidth}
        \centering

        \begin{tabular}{rcc}
\multicolumn{1}{l}{} & Mean & Std. Deviation \\ \hline
\textit{no prediction} & 7.33 & 0.39 \\
\mode{value} & 7.72 & 0.44 \\
\mode{range} & 7.63 & 0.37 \\
\mode{bars} & 7.62 & 0.43 \\
\mode{stars} & 7.61 & 0.35 \\ \hline
\end{tabular}

       \caption{Means and standard deviations}
       \label{tab:PairwiseComparisons:AverageGrade:DescriptiveStats}
    \end{subtable}
    \hfill
    \begin{subtable}[h]{0.61\textwidth}
        \centering

        \begin{tabular}{rccccc}
\multicolumn{1}{l}{} & \textit{no prediction} & \mode{value} & \mode{range} & \mode{bars} & \mode{stars} \\ \hline
\textit{no prediction} & 1 & -.386* & -.298* & -.289* & .279* \\
\mode{value} &  & 1 & .088 & .097 & .107 \\
\mode{range} &  &  & 1 & .009 & .019 \\
\mode{bars} &  &  &  & 1 & .010 \\
\mode{stars} &  &  &  &  & 1 \\ \hline
\end{tabular}
        
        \caption{Post hoc comparisons. Mean differences shown.}
        \label{tab:PairwiseComparisons:AverageGrade:PostHoc}
     \end{subtable}
     
     \caption{Results of the ANOVA for the \textbf{average predicted grade} of the course sets selected by the students. Post hoc comparisons use Bonferroni adjustment. * shows the mean difference is significant at the .05 level.}
     
\label{tab:PairwiseComparisons:AverageGrade}
     
\end{table*}

\begin{table*}[hb]
\centering

    \begin{subtable}[h]{0.37\textwidth}
        \centering

        \begin{tabular}{rcc}
\multicolumn{1}{l}{} & Mean & Std. Deviation \\ \hline
\textit{no prediction} & 4.34 & 0.57 \\
\mode{value} & 3.84 & 0.67 \\
\mode{range} & 3.93 & 0.62 \\
\mode{bars} & 4.02 & 0.56 \\
\mode{stars} & 3.97 & 0.51 \\ \hline
\end{tabular}

       \caption{Means and standard deviations}
       \label{tab:PairwiseComparisons:AverageLoad:DescriptiveStats}
    \end{subtable}
    \hfill
    \begin{subtable}[h]{0.61\textwidth}
        \centering

        \begin{tabular}{rccccc}
\multicolumn{1}{l}{} & \textit{no prediction} & \mode{value} & \mode{range} & \mode{bars} & \mode{stars} \\ \hline
\textit{no prediction} & 1 & 0.502* & 0.413 & 0.313 & 0.369* \\
\mode{value} &  & 1 & -.089 & -.188 & -.133 \\
\mode{range} &  &  & 1 & -.099 & -.044 \\
\mode{bars} &  &  &  & 1 & .056 \\
\mode{stars} &  &  &  &  & 1 \\ \hline
\end{tabular}
        
        \caption{Post hoc comparisons. Mean differences shown.}
        \label{tab:PairwiseComparisons:AverageLoad:PostHoc}
     \end{subtable}
     
     \caption{Results of the ANOVA for the \textbf{average chosen workload} (in hours/week) of the course sets selected by the students. Post hoc comparisons use Bonferroni adjustment. * shows the mean difference is significant at the .05 level.}
     
\label{tab:PairwiseComparisons:AverageLoad}
     
\end{table*}

\subsection{Data Collection and Statistical Tests}

Besides the questionnaires answers, we recorded the set of courses our participants chose in each selection task, as well as their associated grades and workload.
Because this study did not involve interviews or screen recordings, we instrumented \app~to log the consequences of several types of user interactions. These included the partial sets participants progressively built when deciding on their courses, as well as the number of times they opened the explanations of the predicted performances through the \textit{Why} button and the duration of these events.\\





To inquire whether the prediction representations explain the differences in the means of the dependent variables mentioned above (students' decisions and behavior), we carried out two types of statistical tests. We conducted a \textbf{within-subject analysis} with the data of the students exposed to each experimental condition---\textit{specific} or \textit{vague}. For these analyses we used one-way ANOVAs, per dimension of each dependent variable. When the data was found to be not spherical (i.e., the Mauchly’s test failed), we applied a Greenhouse-Geisser correction. All post-hoc tests were corrected for multiple comparisons using Bonferroni corrections. These analyses also included the non-predictive measures obtained under the \textit{no prediction} condition, since all participants were exposed to it.\\ 

The \textbf{between-group analysis} was carried out using a series of t-tests. These analyses consisted of a cross comparison between the measurements of the dependent variables under each representation of the experimental conditions. This yielded a set of 80 comparisons (e.g., average number of selected courses using: \mode{value} and \mode{text}, \mode{value} and \mode{area}, \mode{value} and \mode{color}, \mode{value} and \mode{faces}, and so on).


We also carried out a \textbf{contrast itemset mining analysis}~\cite{discriminative-pattern-mining, Pham2019} to investigate whether some prediction representations may have induced students to select particular groups of courses.

All of our tests were carried out with a significance level $p<0.05$.












\subsection{Results}

We excluded the data of 12 participants from our analyses due to inconsistencies between their answers and the usual enrollment habits of \college~students\footnote{These 12 participants chose either more than 6 courses or less than 3. The first scenario is not allowed at \college. On the other hand, enrolling in less than 3 courses mostly happens under very specific circumstances (e.g., when a student is at their very last academic term). Hence, our participants did not have any valid reason to choose so few courses.}. Our analyses are thus based on the data of 79 participants---37 who were exposed to the \textit{vague} representations and 42 who used \app~under the \textit{specific} ones.

We present our findings along three axes, namely the students' decisions, their behaviors, and their preferences.

\subsubsection{Students' Decisions}


The decision of a student after a course selection task with \app~is defined by the set of courses selected. We elaborate on our findings in three stages. In the first stage, we discuss the results of the within-subject and between-group analyses.
In a second stage, we compare the chosen courses with all the partial sets ever composed by the students. This analysis aims at detecting the grade maximization effect observed in Study 1. In a third and final stage, 
we report the results of an analysis based on contrast itemset mining~\cite{discriminative-pattern-mining} on the courses chosen by the students. The goal of this analysis is to identify groups of courses that are preferred by the participants exposed to a particular type of visual representation.

\paragraph{Within-subject analysis.} 
Our analysis did not yield any significant differences within the \textit{vague} representations condition for any of the dimensions of our dependent variable. On the contrary, we found significant differences between the \textbf{average predicted grade} (F(3.028, 124.146) = 8.097, p = 0.0005, $\eta_{p}^{2}=0.165$) within the students exposed to the \textit{specific} representations condition. The post hoc tests revealed significant differences in the means of this dimension between the variant without prediction and each of the \textit{specific} prediction representations (see Table~\ref{tab:PairwiseComparisons:AverageGrade:DescriptiveStats} and~\ref{tab:PairwiseComparisons:AverageGrade:PostHoc}). Note that the \textit{no prediction} condition reaches a lower mean in the average predicted grade than those in the \textit{specific} representations condition.




\begin{figure*}[hb!]
    \centering
    \begin{subfigure}[t]{0.235\textwidth}
        \centering
        \includegraphics[width=0.85\textwidth]{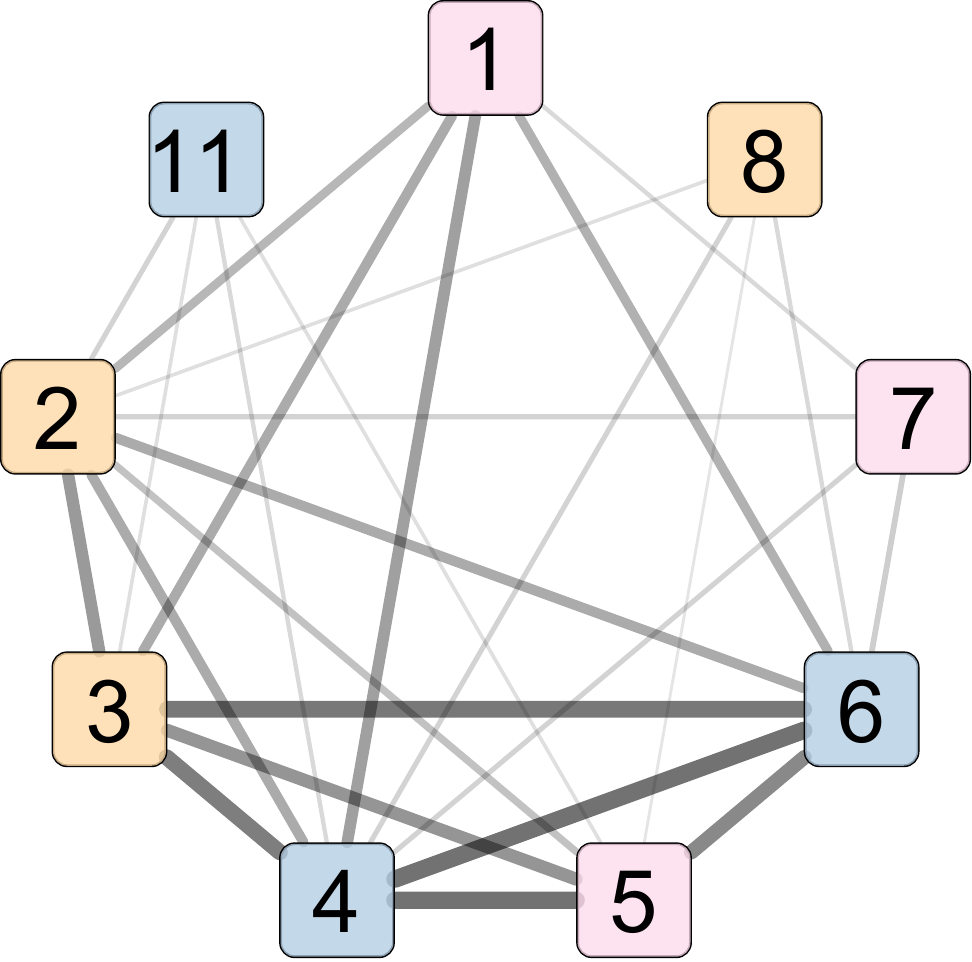}
        \caption{No prediction - \textit{Specific}}
        \label{figure:graph:noPredictionSpecific}
    \end{subfigure}%
    ~
    \begin{subfigure}[t]{0.235\textwidth}
        \centering
        \includegraphics[width=0.85\textwidth]{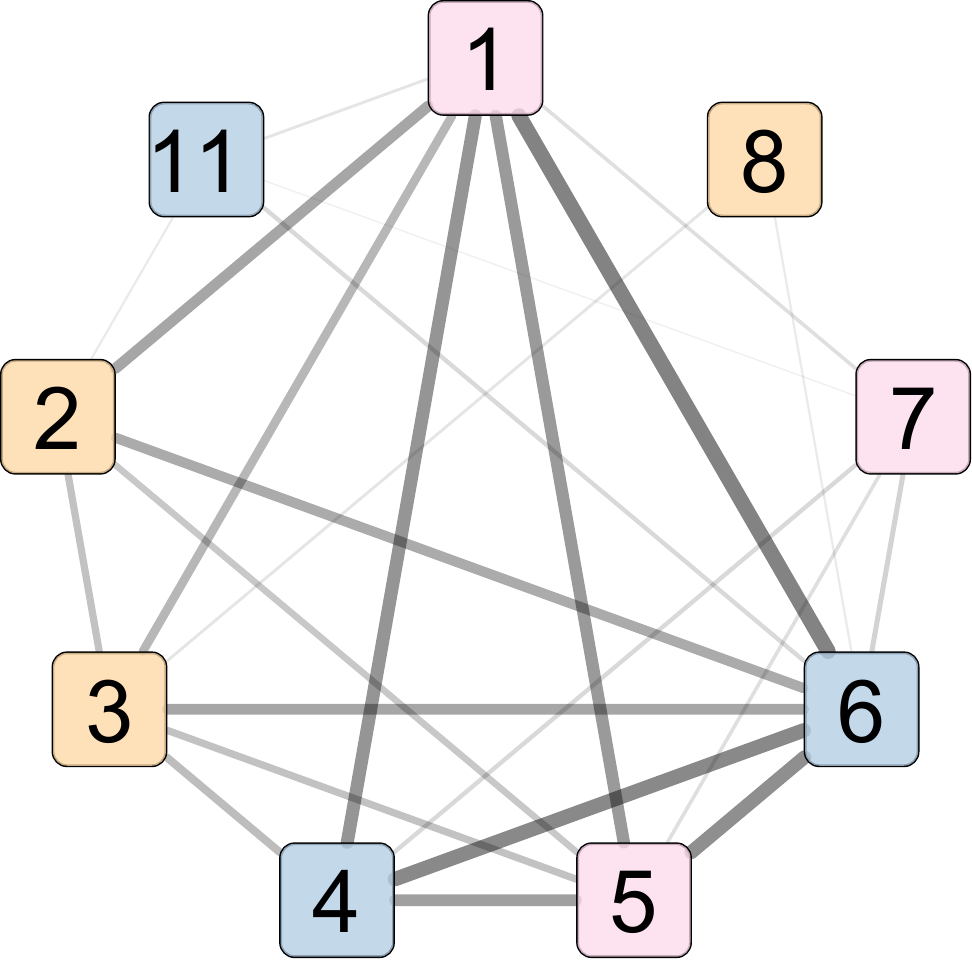}
        \caption{No prediction - \textit{Vague}}
        \label{figure:graph:noPredictionVagues}
    \end{subfigure}%
    \\[0.2in]
    \begin{subfigure}[t]{0.235\textwidth}
        \centering
        \includegraphics[width=0.85\textwidth]{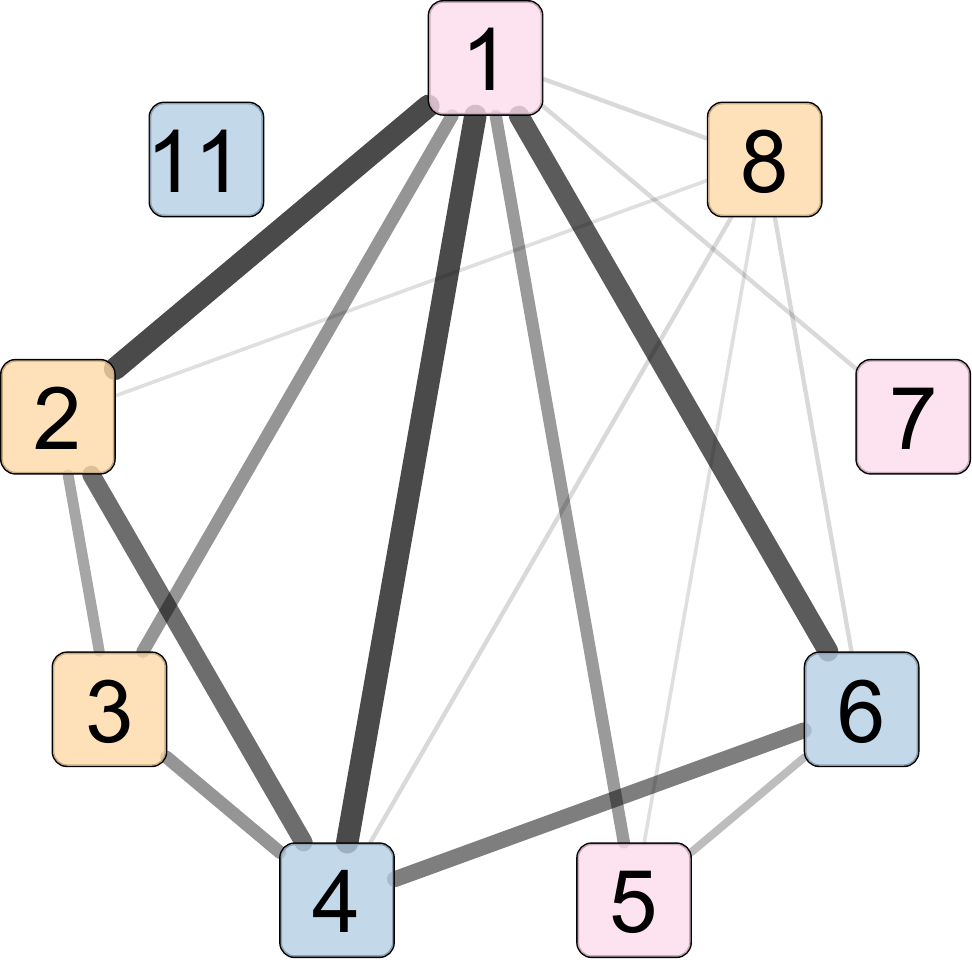}
        \caption{Value}
        \label{figure:graph:value}
    \end{subfigure}%
    ~
    \begin{subfigure}[t]{0.235\textwidth}
        \centering
        \includegraphics[width=0.85\textwidth]{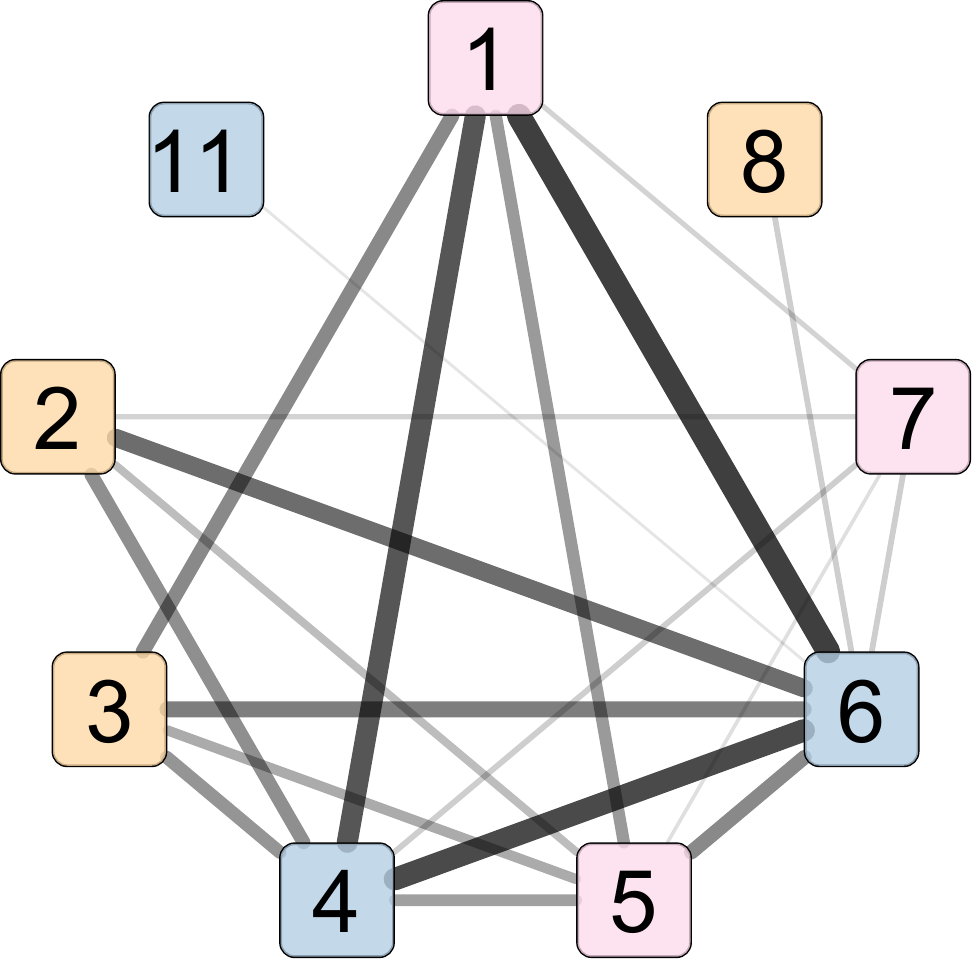}
        \caption{Range}
        \label{figure:graph:range}
    \end{subfigure}%
    ~
    \begin{subfigure}[t]{0.235\textwidth}
        \centering
        \includegraphics[width=0.85\textwidth]{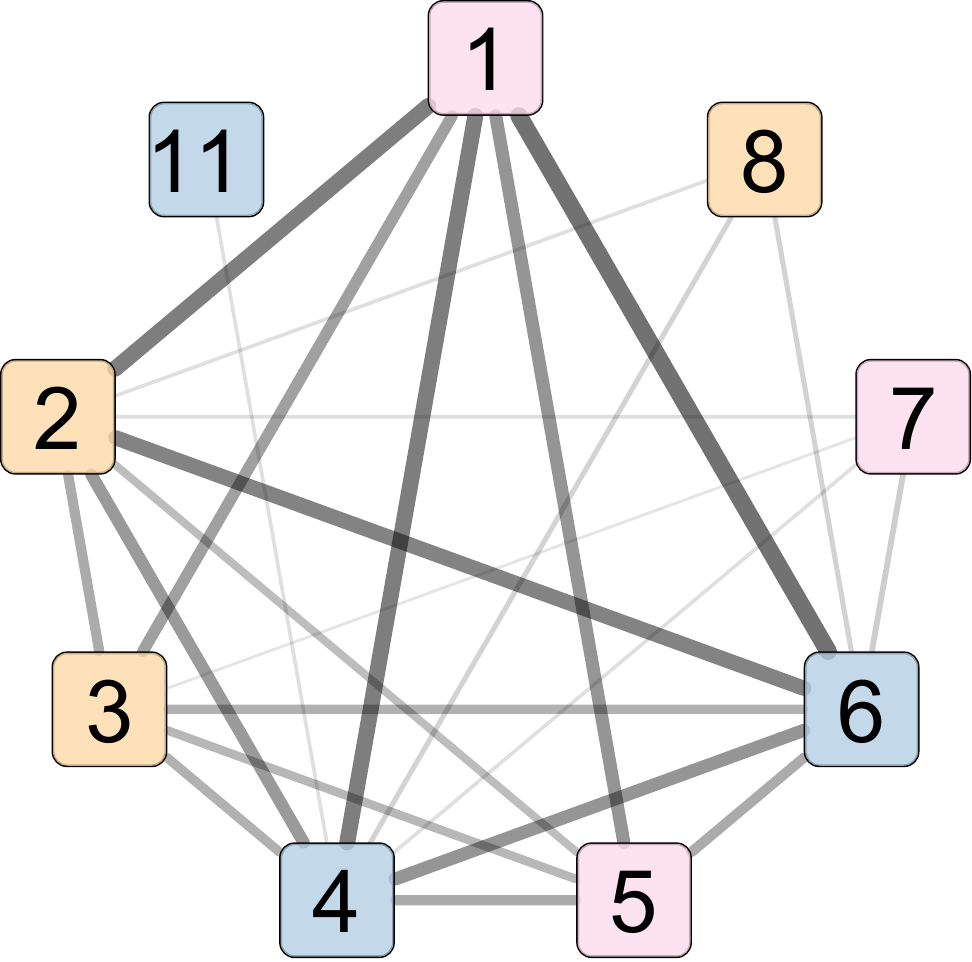}
         \caption{Bars}
        \label{figure:graph:bars}
    \end{subfigure}%
    ~
    \begin{subfigure}[t]{0.235\textwidth}
        \centering
        \includegraphics[width=0.85\textwidth]{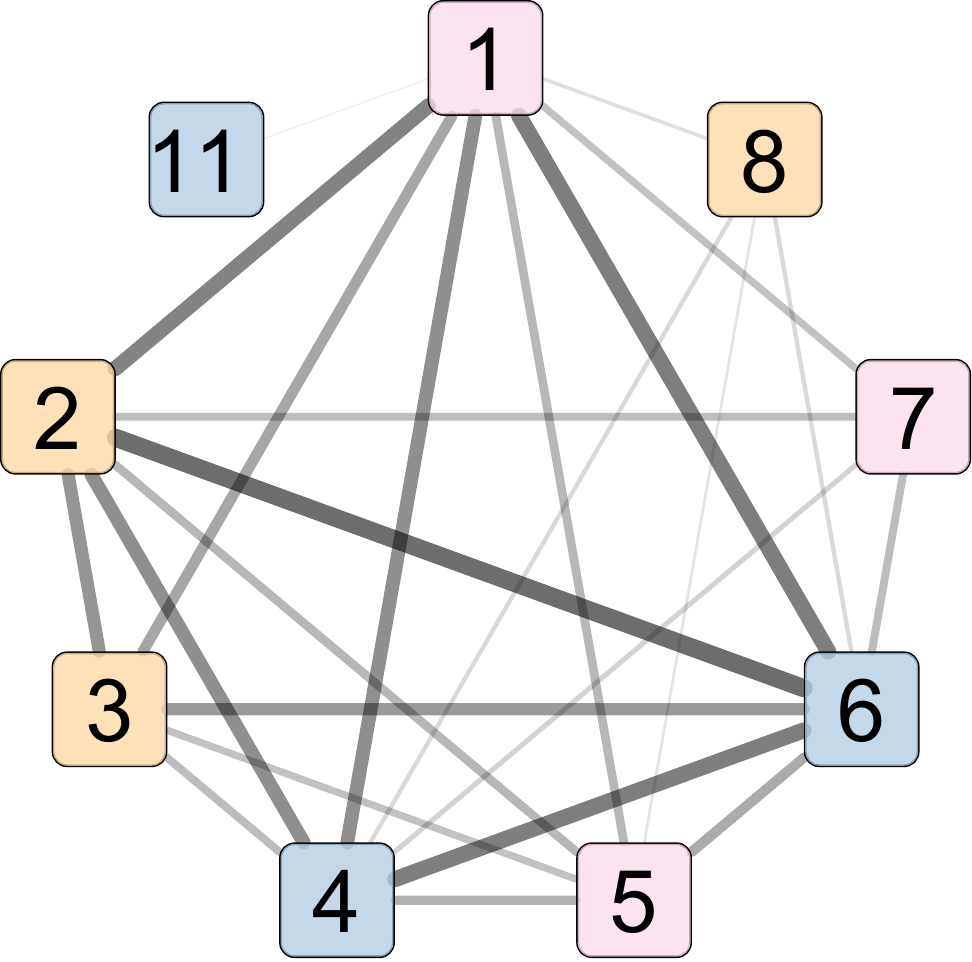}
         \caption{Stars}
        \label{figure:graph:stars}
    \end{subfigure}%
    \\[0.2in]
    \begin{subfigure}[t]{0.235\textwidth}
        \centering
        \includegraphics[width=0.85\textwidth]{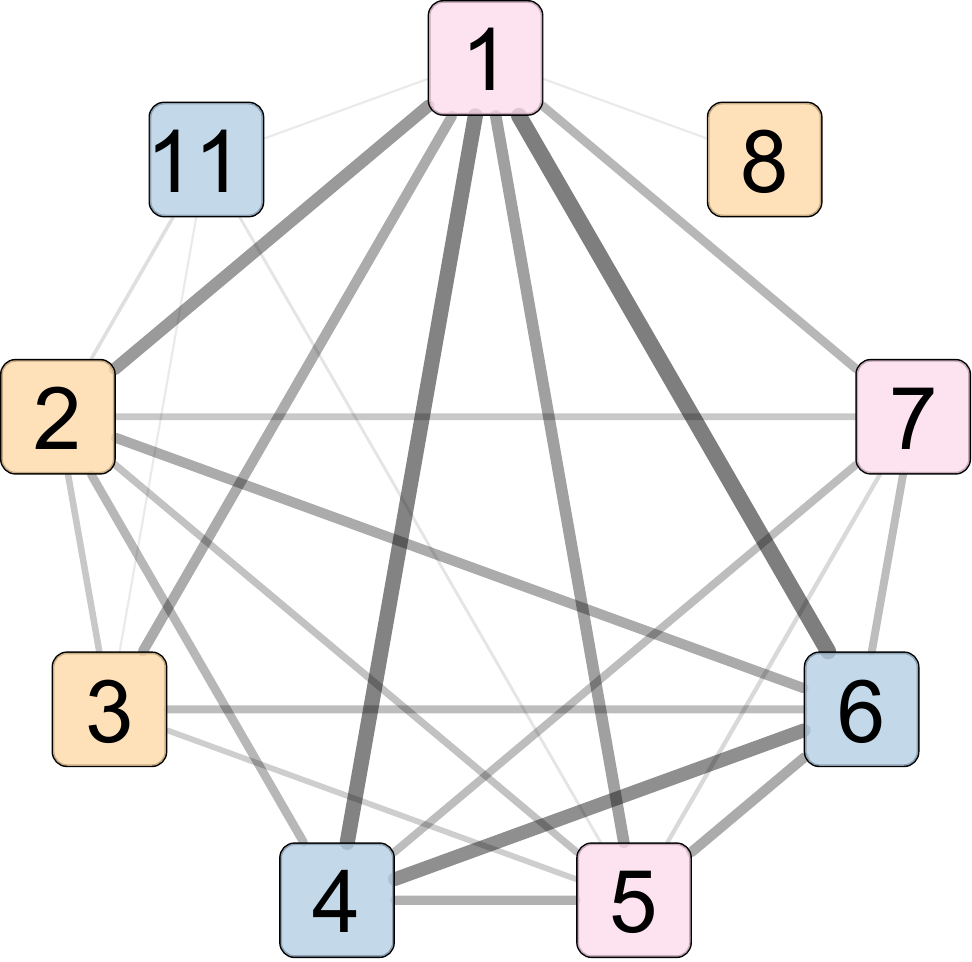}
         \caption{Area}
        \label{figure:graph:area}
    \end{subfigure}%
    ~
    \begin{subfigure}[t]{0.235\textwidth}
        \centering
        \includegraphics[width=0.85\textwidth]{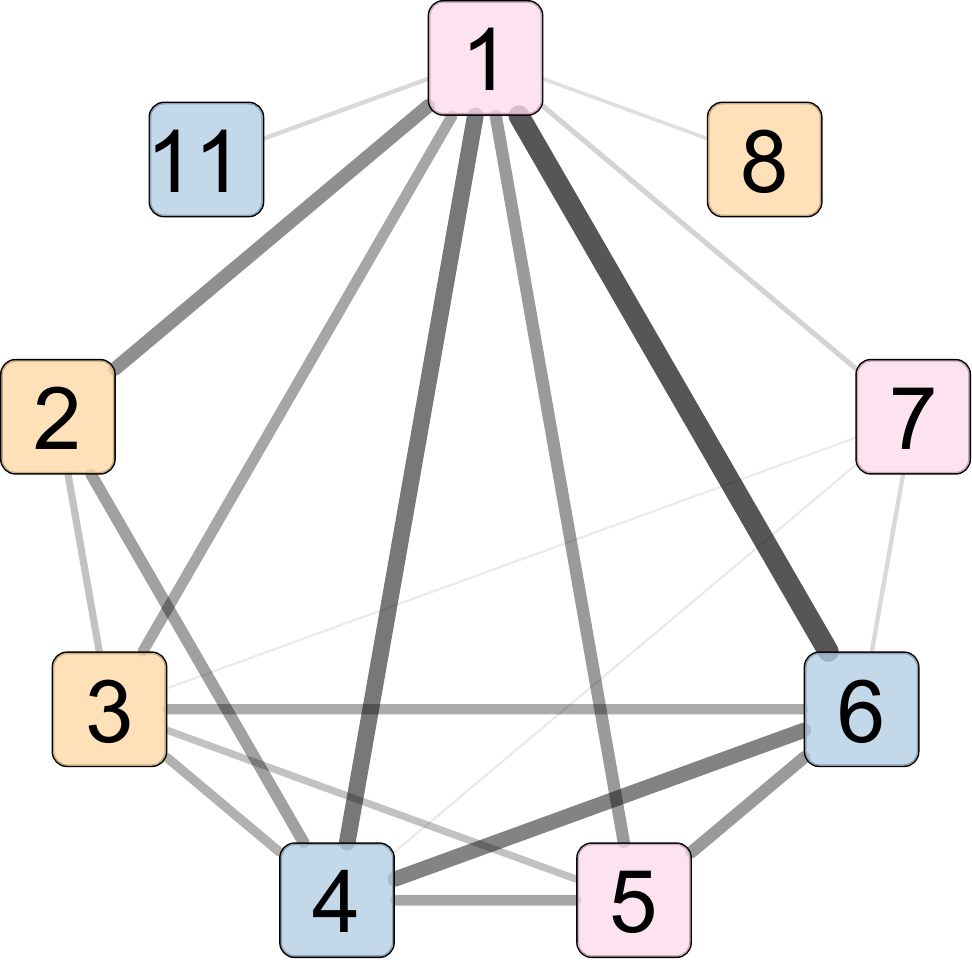}
         \caption{Color}
        \label{figure:graph:color}
    \end{subfigure}%
    ~
    \begin{subfigure}[t]{0.235\textwidth}
        \centering
        \includegraphics[width=0.85\textwidth]{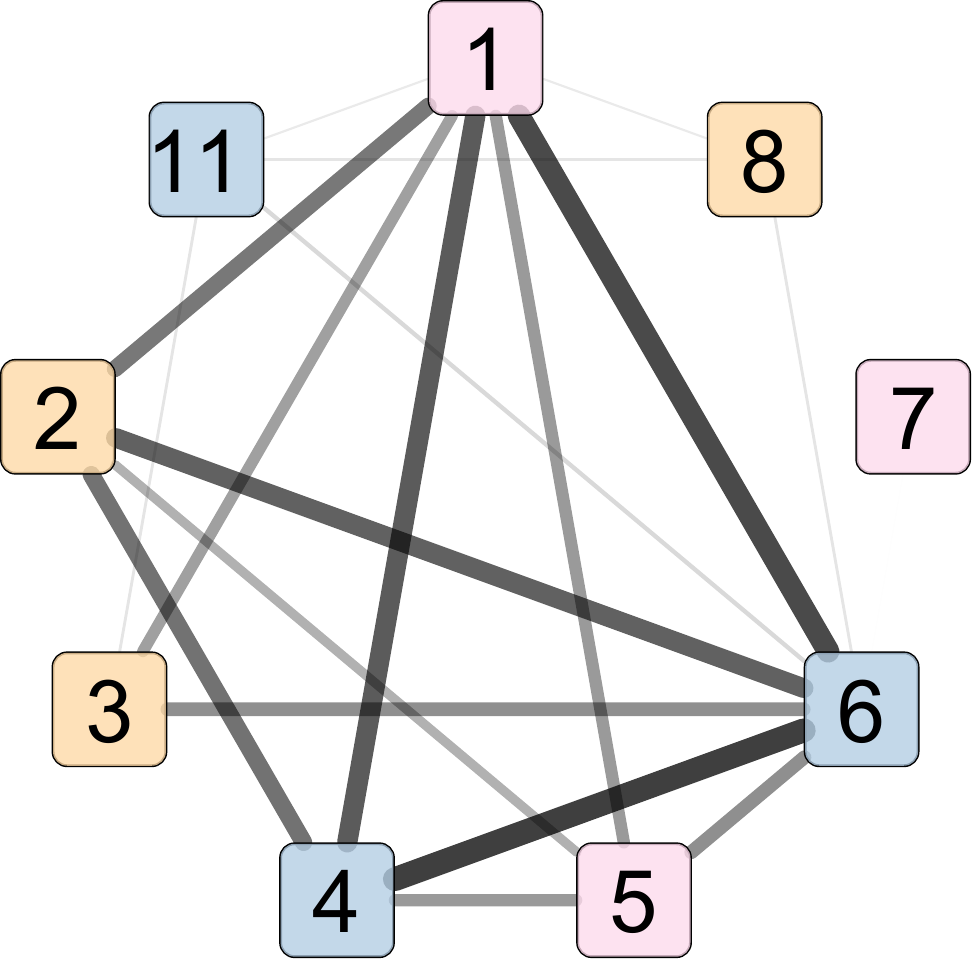}
         \caption{Text}
        \label{figure:graph:text}
    \end{subfigure}%
     ~
    \begin{subfigure}[t]{0.235\textwidth}
        \centering
        \includegraphics[width=0.85\textwidth]{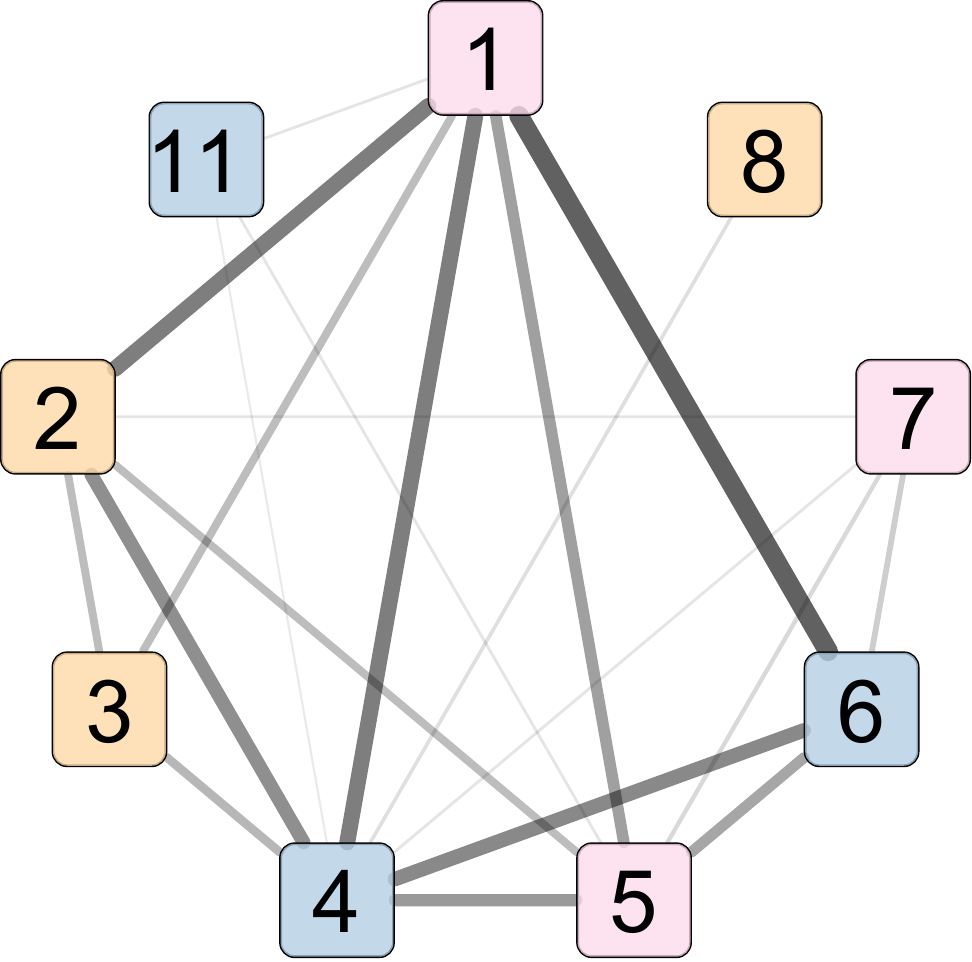}
         \caption{Faces}
        \label{figure:graph:faces}
    \end{subfigure}
    \caption{Co-occurrence graphs of the courses students selected when exposed to different types of prediction representations.}
    \label{figure:graphs}
    \Description{Several graphs composed by circular nodes and connected through solid lines show the co-ocurrence of courses in the sets selected by the students. Thicker edges indicate more frequent co-ocurrences. Courses appear colored according to their type.}
\end{figure*}

We also detected significant differences between the means for the \textbf{average chosen workload} (F(2.905, 119.102) = 5.771, p =0.001, $\eta_{p}^{2}=0.123$). The pairwise comparisons revealed differences between the \textit{no prediction} condition and the \mode{value} and \mode{stars} representations (see Tables~\ref{tab:PairwiseComparisons:AverageLoad:DescriptiveStats} and~\ref{tab:PairwiseComparisons:AverageLoad:PostHoc}). Note again, that the mean for the \textit{no prediction} condition is higher than the previously highlighted representations.





\paragraph{Between-group analysis.} We carried out a set of t-tests between groups of students exposed to both the \textit{specific} and \textit{vague} representations. This round of experiments yielded a significant difference in the \textbf{average number of selected courses} for the \mode{text} (\textit{m} = 5.225) and \mode{bars} (\textit{m} = 4.38) representations, \textit{t}(80) = 2.09, \textit{p} = 0.04.

\paragraph{Grade Maximization Effect.} \label{subsubsec:students_objectives} As done for Study 1 (Section~\ref{subsubsec:influence_decisions}), we looked at the maximal grades of all the course sets ever composed by a student during an interaction with~\app. We then calculated, for each visual representation, the average percentile of the maximal grade of the selected course. The average percentile ranges from 53.37 (standard deviation $\sigma=29.38$) for the \mode{stars} representation, to 66.44 ($\sigma=24.30$) for the  \textit{no prediction} mode. The average percentile across all interactions is 59.42 ($\sigma=26.64$). The trends are similar for the average grade of the course sets, whereas for the minimal grade the highest average percentile is 32.78 ($\sigma=22.70$). These results contest the grade maximization effects we observed for the GPA in Study 1.

\paragraph{Itemset Mining on Courses.} 
We also investigated whether some prediction representations may have leaned students towards choosing specific courses. For this purpose, we looked at the co-occurrence graphs of the courses the students selected per visual representation (Figure~\ref{figure:graphs}). The nodes of these graphs represent the courses available for enrollment in the study's selection tasks. Thicker edges denote higher co-occurrence. 
We observe some recurrent cliques in all scenarios, e.g., the set \{ 1, 2, 4, 6 \} is prominent in almost all cases. Motivated by this insight, we carried out a deeper analysis based on contrast itemset mining~\cite{discriminative-pattern-mining, Pham2019}. 
This technique finds groups of courses that co-occur more frequently in a visual representation than in others. We measure the relevance of those groups via the \emph{growth ratio} score~\cite{discriminative-pattern-mining}, which given two categories, defines the ratio of the frequencies\footnote{That is, the number of course sets that contain the group divided by the total number of course sets in the category. We considered groups of courses occurring in at least 10 course sets.} of a group of courses in each of the two categories. Values larger than 1 denote \quotes{interesting} groups.\\


\begin{figure*}[hb!]
  \centering
  \begin{subfigure}[b]{0.5\linewidth}
    \centering\includegraphics[width=129pt]{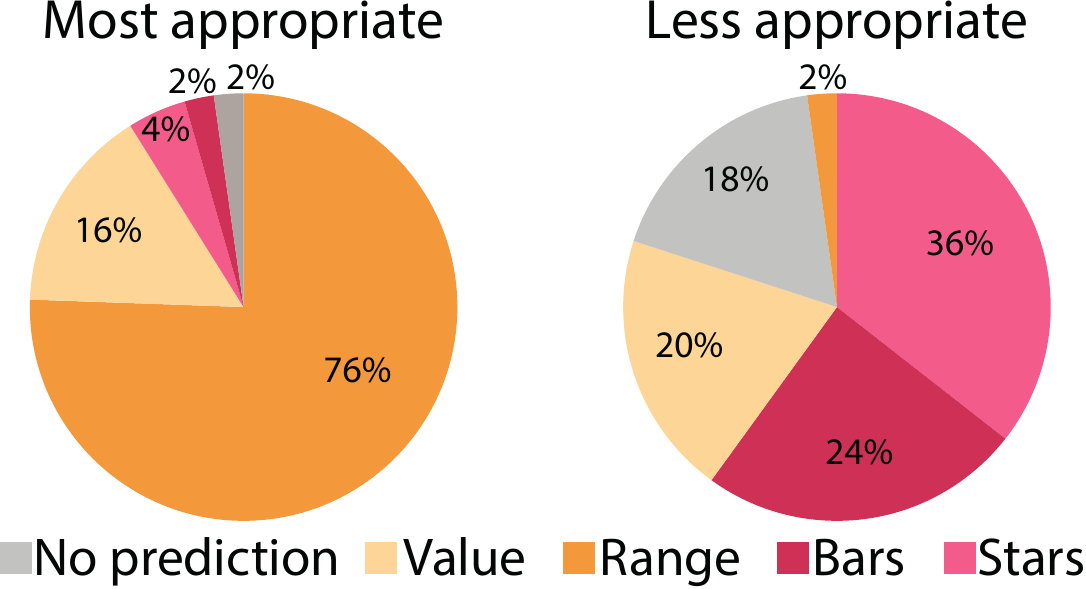}
    \caption{\label{fig:appropriatness:specific}Specific}
  \end{subfigure}%
  \begin{subfigure}[b]{0.5\linewidth}
    \centering\includegraphics[width=129pt]{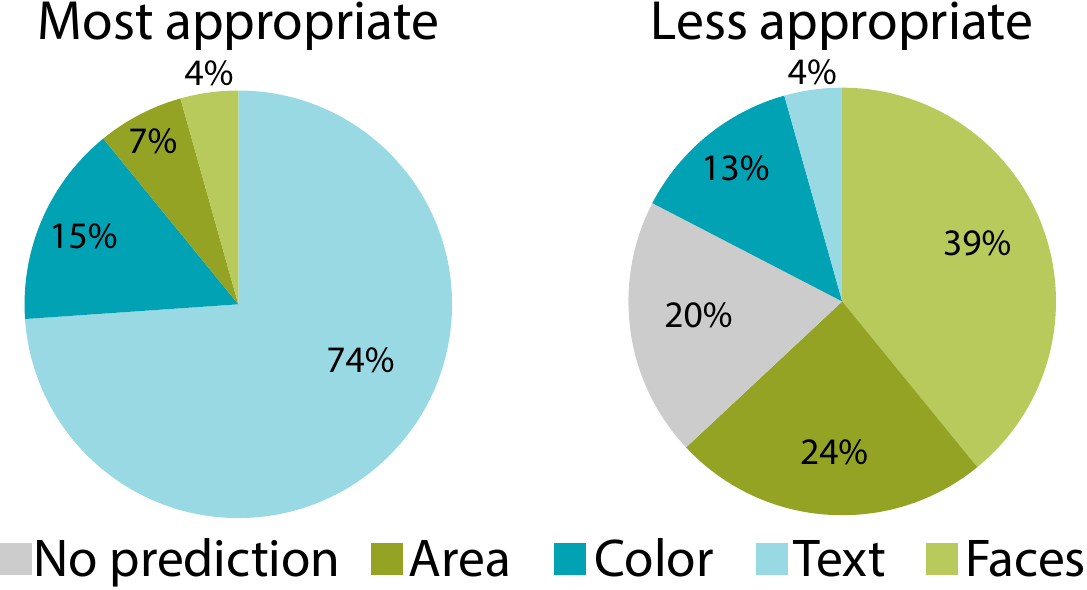}
    \caption{\label{fig:appropriatness:vague}Vague}
  \end{subfigure}
  \caption{Most and least appropriate prediction representations according to the participants' opinions.}
    \label{fig:appropriatness}
    \Description{Four pie charts show the percentages of people who considered the visual representations used in Study 2 the most and least appropriate. Each sector of the pie chart is colored differently to depict the type of visual representation.}
\end{figure*}

Albeit frequent everywhere, the course group \{ 1, 2, 4, 6 \} is 4.73 times more frequent---with 95\% confidence interval (CI) $(2.85, 7.81)$---in the  \textit{no prediction} scenario than in the scenario with the \mode{text} representation. 
Similar scores can be found between the scenario with \textit{no prediction} and the \mode{faces} visual representation---growth ratio 3.68 with CI $(2.15, 6.34)$.
The set \{ 1, 2, 5, 6 \} is prominent in the scenario without prediction as it is 5.25 times more frequent---CI $(1.78, 15.46)$---than for the \mode{text} representation. 
Conversely the group of courses \{ 3, 4, 5, 6 \} is 2.86 times more frequent---with CI $(1.67, 4.92)$---in the \mode{bars} and \mode{faces} representations than in the scenario with \textit{no prediction}. The growth ratio scores were calculated on sets of courses selected by disjoint groups of students.






\subsubsection{Process}
\label{subsubsec:results_process}
We present the results in regards to this variable in line with our within-subject and between-group analysis protocols.
We did not find significant differences for the dimensions of behavior in our within-subject analysis. In regards to the between-group analysis, our t-tests revealed significant differences in \textbf{the average number of times the users opened the explanations} for the (i) the \mode{text} (\textit{m}=4.78) vs. the \mode{value} representations (\textit{m}=2.4, t(51) = 2.44, $p=0.02$); (ii) \mode{text} (\textit{m}=4.78) vs. \mode{bars} (\textit{m}=2, t(53) = 3.04, $p=0.004$); (iii) \mode{faces} (\textit{m}=4.04) vs. \mode{value} (\textit{m}=2.98, t(42) = 2.13, $p=0.04$); and (iv) \mode{faces} (\textit{m}=4.04) vs. \mode{bars} (\textit{m}=2.28, t(46) = 2.94, $p=0.006$).  






\subsubsection{Preferences}

We asked students, in the closing questionnaire, their opinion on the most and least appropriate prediction representations. As summarized in Figure~\ref{fig:appropriatness}, there seems to be a consensus when it comes to their preferred prediction representations: 76\% of the students favored the \mode{range} within the \textit{specific} types, whereas 74\% of the students deemed the \mode{text} the most suitable to convey predictions among the \textit{vague} representations.








\paragraph{Preferred Representations}

When explaining their preferences, most participants who favored the \mode{range} representation compared it with the \mode{value} one and highlighted the capacity of the former to convey uncertainty. This was often mentioned as a booster of their trust in the grades predicted by the tool: \quotes{\textit{A range implies that the prediction is subject to uncertainty and it gives a more realistic view than a specific grade. It seems to me a better approach to show the predictions. It seemed more credible and gave me more information than the other options.}}~[P40S]; \quotes{\textit{I find it difficult to believe that I will get exactly the grade shown by the exact value. However, a range seems more credible to me.}}~[P32S]; \quotes{\textit{It is better for students to see a range of possible values for their grades, since this indicates how much our grades may vary if we do not keep our effort level; this message is impossible to convey with an exact value.}}~[P10S]; \quotes{\textit{I consider it more reliable, as it shows a margin in which my grade will be located. I can consider that at least I have a margin of error [...] Compared to the other ones, despite being very visual, they don't tell me much at the end of the day.}}~[P05S]; \quotes{\textit{It is better to have a range, I can't trust an exact grade.}}~[P24S]; \quotes{\textit{It gives us more confidence because we know that a prediction has a margin of error.}}~[P37S]; \quotes{\textit{I feel that with an exact grade there is more possibility of error. Instead, a range lets me know, more or less, the grade I may get.}}~[P02S].


The students who used \app~with the \textit{vague} representations preferred the \mode{text} mainly because of its simplicity and directness: \quotes{\textit{It is much clearer and more explanatory.}}~[P03V]; \quotes{\textit{It is simple and concise.}}~[P09V];  \quotes{\textit{I prefer to be presented with things in a more direct way, and this message is, to some extent, encouraging.}}~[P10V]; \quotes{\textit{It is faster to grasp.}}~[P17V]. Students also highlighted that, compared to the others from the \textit{vague} category, the \mode{text} representation does not require interpretation: \quotes{\textit{It is easier to understand, it leaves nothing to interpretation.}}~[P45V]; \quotes{\textit{It gives me an answer that is easy to understand. With the colors or the faces, I have to infer what the symbols are and what each means.}}~[P11V]; \quotes{\textit{It seems the most appropriate to me because no previous explanation is needed to understand how it works. It is intuitive and it tells directly how I would perform in a course.}}~[P30V]. Other students commented on how close the textual messages were to the advice provided by their advisors or other students:
\quotes{\textit{I felt that in a certain way, it encouraged me to take the courses, because the language used is similar to the one a friend from my degree would have used when talking about the courses.}}~[P28V]; \quotes{\textit{The information is somewhat similar to the recommendations my advisor would give me in person.}}~[P42V].

\paragraph{Non-preferred Representations}

The opinions about the least appropriate representations were varied. The \mode{stars} and \mode{faces} stand out as the least preferred representations according to 36\% and 39\% of the participants. They are followed by the \mode{area} and the \mode{bars}, both appearing in 24\% of the answers.

In the \textit{specific} category, the \mode{stars} and the \mode{bars} were deemed as not precise enough, distracting, and even \quotes{not serious} for an academic context. The comments for these representations also highlighted the lack of a numeric representation of the predicted grades, as illustrated by this exemplary statement: \quotes{\textit{It does not provide much feedback. What I want to see is my grade.}}~[P02V].

The \mode{faces} were rejected because it was hard for students to discern differences between the representations shown. It was common for students to state that it was hard to distinguish the degree of happiness or sadness in the facial expressions. A similar problem was reported about the circular marks used in the \mode{area} representation. The sizes of the circles were considered not easily distinguishable.

\section{Discussion}

Our discussion is initially structured along the three same axes used to present our experimental results in light of our research questions \textbf{RQ1} and \textbf{RQ2} (Sections~\ref{subsec:discussion_decisions}--\ref{subsec:discussion_preferences}). Additionally,  in Section~\ref{subsec:ethical_considerations}, we discuss the ethical considerations of using GPA-based predictions for course selection and recommendation. 


\subsection{Students' Decisions} \label{subsec:discussion_decisions}
Study 1 tackles \textbf{RQ1} by investigating the effect on students of displaying performance predictions during term planning. The observations of this study suggest that predictions can make students embrace a grade and GPA optimization approach, disregarding other important factors (Section~\ref{subsubsec:influence_decisions}). Study 2 inquired whether this behavior was caused by the mere presence of individual course performance predictions, and whether those predictions induce framing effects on the students (\textbf{RQ2}). The results of this second study (Section~\ref{subsubsec:students_objectives}) suggest that, at least for the individual grades, predictions per se do not induce a grade maximization effect. We remark, however, that Study 2 left out the prediction of the GPA. This raises the question of whether this factor might have been the trigger of the maximization effect observed in Study 1, as it has a greater impact on the career of the students than the individual course grades of an academic term.


Study 2 also suggests that the students' decision process and their final choices are indeed influenced by the type of prediction representation (\textbf{RQ2}). While no visual representation seems to have favored a grade maximization effect, the \textit{specific} representations seemed to have leaned students towards more optimistic predictions and lighter workloads (see Tables~\ref{tab:PairwiseComparisons:AverageGrade:PostHoc} and \ref{tab:PairwiseComparisons:AverageLoad:PostHoc}). This was not the case for the visualizations located at the \textit{vague} end of our spectrum of prediction representations. All this suggests that when exposed to \quotes{countable} predictions, students put more effort on the course selection task. This is confirmed by the fact that, on average, the students composed more partial sets when exposed to the \textit{specific} representations: The \mode{value} and range representations lead the way with the longest sequence of interactions---8.02 and 7.73 sets on average---before making a decision (the total average is 6.59). This indicates that \textit{specific} (i.e., countable) visual representations make students iterate more over their enrollment options, which could be a sign of a deeper and more critical reflection process.

If the courses chosen by the students do not point to a grade maximization effect, then the students must be also taking the workload into account. This assertion is suggested by our itemset mining analysis. The group of courses \{ 1, 2, 4, 6 \}, which is prevalent in all the levels of our experimental conditions, includes 3 courses with a workload of 3 hours each, and one course with a workload of 5 hours. This \quotes{formula} actually corresponds to the lightest possible combination of courses in terms of workload. Nonetheless, this logic does not apply to all popular groups of courses. For instance, the itemset \{ 3, 4, 5, 6 \}---prevalent mostly in the \textit{vague} representations \mode{bars} and \mode{faces}---leads to a high workload. Conversely, the popularity of this set can be explained by the location of its components in the visualization of the academic program: These appeared together at the left-most end of the upper row of courses available for enrollment. A similar observation applies to the group of contiguous courses \{ 1, 2, 5, 6 \}, particularly prominent in the \textit{no prediction} scenario. Altogether, this evidence suggests that the students do care about workload when no prediction about their GPA is shown, and that the \textit{vague} representations may induce students to think less over their enrollment choices---an overreliance effect. Furthermore, we highlight a preference for courses in the upper row of available courses in the program. This conforms to the strategy expressed by some students in Study 1 and also confirmed by the demographics questionnaire of Study 2: when deciding on their enrollment, students favor courses located at the level of the upcoming semester in their study program.

\subsection{Course Selection Process} 
We discuss this axis in terms of two aspects of the course selection process, namely, the strategy used by the students and their interactions with \app's explanations. We also discuss the potential risk of overreliance and automation-complacency effects.

\paragraph{Strategy.} Our analysis of the students' interactions with \app{} showed a recurrent two-stage strategy to compose set of courses. In the first stage, students generally added three courses to the grades prediction panel. This was followed by an exploration phase in which they added and removed courses repeatedly. This behavior was common regardless of the prediction visual representation (even in the \textit{no prediction} scenario). While this might imply students picked up and removed courses driven by some sort of optimization objective, our analysis of Section~\ref{subsubsec:students_objectives} indicates that they did not necessarily settle for the most optimistic predictions. 
The lower average workloads observed for the \textit{specific} visual representations reiterates the role of the workload in the students' approach. 

\paragraph{Interactions with the Explanations.} In Study 1, our participants exhibited a weak interest on the explanations \app~provided, both for the course difficulty estimators and the performance predictions (accessible via the \emph{Why} button). Nevertheless, we obtained hints about possible causes of that lack of interest. Some of the students argued that the explanations were too long and far from obvious, which is consistent with the observation that the first interaction was comparatively long (53 seconds on average) and was rarely followed by a second one. When designing our second study, we took action in this regard and simplified the performance explanations. This, however, did not increase the interest of the students: In 49\% of the course selection tasks, our participants did not interact with the performance explanations at all, and only in 11\% of the cases there was more than one interaction. The total time invested in reading the explanations was on average 14 seconds, although we expect it to be shorter than for Study 1 since the explanations were more concise. The within-study and between-group analyses described in Section~\ref{subsubsec:results_process} did not show any significant difference in the total interaction time with the explanations as a consequence of the visual representation of the prediction. However, the \textit{vague} representations \mode{color} and \mode{faces} led to significantly more interactions than the \textit{specific} visualizations \mode{value} and \mode{bars}. These higher numbers of interactions with \app's explanations could be explained by the interpretation overhead incurred by the \textit{vague} representations. Nevertheless, explanations might have been less needed for the \textit{specific} representations, as these make the grades predicted more directly readable.

\paragraph{Overreliance and Potential Complacency Effects.} The lack of interest in the explanations provided by \app~might suggest some sort of automation complacency~\cite{goddard2012automation} in the students regarding \app's performance predictions. Although it could be said that students did not need major explanations---because they trusted the system---our analyses rather suggest that they were, in most of the cases, not very interested in understanding what was happening inside the tool. Explanations incur, however, a cognitive load on users. A promising research direction could be to decide the right stages of the course selection process where explanations are pertinent and desirable. 
An exciting venue for future research would be the exploration of student-generated explanations of their performance. This type of explanation design has shown benefits in the visualization of complex scientific phenomena~\cite{ryoo2014designing}.

The evidence we gathered on overreliance and potential complacency effects, however, is not conclusive. Further investigations are needed in this area to better understand and fully characterize these effects.


\subsection{Preferences} 
\label{subsec:discussion_preferences}
The preferences of the students elicited via our studies indicate that they value two attributes in performance predictions, namely \emph{credibility} and \emph{directness}. The first factor is corroborated by their preference for the \mode{range} (overall, the most preferred \textit{specific} representation) over the \mode{value}. Indeed, students rated predictions with ranges as more reliable and credible than exact values, as ranges convey more information. 
These observations are consistent with existing studies of the cognitive preferences of people regarding AI agents~\cite{Frnkranz2019OnCP}. The evidence suggests that, from the perspective of user acceptance, the plausibility of a prediction or explanation, i.e., its concordance with the users' background and common sense, is as important as its comprehensibility or simplicity.
This credibility dimension may also explain the preference of students for the \mode{text} over the other \textit{vague} representations, although this preference can also be explained by the directness of textual predictions. This was explicitly stated by several participants of Study 2, who valued textual predictions as direct, simple, and easy to understand. However, it is equally plausible that textual recommendations generated more trust because they expressed messages that were close to what a human advisor or colleague would say.

\subsection{Ethical Considerations of Grade- and GPA-based Predictions} \label{subsec:ethical_considerations}
An important ethical concern of our work arises from profiling students based on their course grades and GPA. Such predictions have been discussed as a potential threat to the students' potential and self-efficacy~\cite{beattie2014creepy, greller2012translating} and must be tempered with caution. As we mentioned earlier, \app~does not seek to replace the human advisors. Rather, it intends to support and facilitate the student-advisor dialogue through a data-driven approach. There exist factors outside the student's academic environment that performance-based predictive models cannot account for (e.g., extracurricular activities, family and health issues). Thus, we highlight the need for human judgment on top of \textit{any} data-based academic performance prediction. Our observations suggest that the decisions and recommendations derived from \app---and similar tools---must remain on the human side of the academic advising process. This is a key aspect of interactive visualization technologies where \quotes{humans in the loop} make decisions and perform analytical tasks based on data. That being said, it is also important to highlight that even with human intervention, overreliance effects may arise. Moreover, students ultimately decide their enrollment based on factors that may change after their advising meetings (e.g., availability of places in courses, scheduling constraints). These aspects are beyond the advisors' reach and are often handled exclusively by the students, at the exact moment of enrollment. Thus, more than recommending which courses could be taken, advisors should provide students with guidelines on the criteria to consider when deciding on their enrollment. Performance and workload should not be the only factors to observe, and other considerations will likely include the specific context of each HEI and the personal circumstances of each student.\\

We also highlight the reductionist nature of the GPA in describing the students’ performance. After all, the learning process comprises other aspects---and, hence, other metrics (e.g., development of learning outcomes, levels of engagement, the students’ learning style, or teaching preferences)---that might be more suitable for prediction. Such metrics, however, are rarely systematically collected by HEIs. When measured, they are often kept by instructors to reflect on specific, localized activities. Therefore, they seldom become part of the students’ official academic record. At \college, for example, the GPA is the official performance metric that students are exposed to throughout their career. It was also the only performance indicator readily available for prediction. This practical limitation forced us to study \app~with a focus on GPA.


One promising future perspective of this work is to elicit conversations with policymakers on alternative performance metrics that could be gathered at HEIs to further empower students and counselors in their use of educational data. Learning strategies are increasingly more oriented to emphasize learning through the demonstration of what a student is able to do with the knowledge they acquire or develop
~\cite{Jonassen1994Thinking}. Moreover, several studies suggest that how students are assessed impacts their learning performance~\cite{schwartz1997problem, oh2003assessment, QAACode}. Therefore, assessment should also be focused on measuring the learning quality, rather than the learning quantity~\cite{Struyven2006}. To stress these aspects,  alternative assessment activities require students to demonstrate of thinking and problem-solving skills, involvement or engagement, performing a significant task, creating an artifact or product, etc. They also resort to portfolios, case-based or peer assessments, and observation of students group process ~\cite{Struyven2006, Maclellan2004}. 

Given this variety of assessment activities, there is a myriad of indicators that could be used as proxies of student performance e.g., level of engagement in meaningful activities, quality of the interactions between peers in collaborative tasks, reflections about students' learning during a design/creation process, certificate or badge achievements, student outcomes observation when doing an activity or working in groups. The increasing penetration of LMS, MOOCs, and learning apps may enable monitoring these indicators and include them as part of the data students and advisors could visualize and discuss during their meetings. However, until enough data on alternative metrics is available, our findings should be interpreted considering the limitations of the GPA discussed throughout the paper.



\section{Limitations and Open Questions}

An obvious threat to the ecological validity of our studies is the use of a fictional academic history and courses from external CS curricula. We acknowledge that the stakes are higher in real-world scenarios, where poor enrollment decisions have a real impact on the students' life. 
That being said, we did not find any indication that introducing courses from other curricula in our studies got in the way of our participants' decisions. 
Although it could be argued that these decisions might have been made without much consideration,
our video analysis of the data from Study 1 showed that students indeed engaged in the course selection tasks, often thinking aloud about their enrollment options. The setting of Study 2 did not allow for the collection of video data. However, our quantitative analyses were based on the data of participants whose decisions complied with the usual enrollment patterns of \college~students. The use of synthetic data also allowed us to reduce the \textit{vox populi} effect, toward a more objective course selection process.

It is important to remark that course selection is also affected by non-academic aspects. Extracurricular workload, health and family issues, and many other factors play a role in the decisions and academic performance of students~\cite{KARDAN20131}. In this regard, our results should be taken with a grain of salt. In the same vein, the evidence gathered through our studies is not sufficient to rule out the presence of extra-representational factors (e.g., preferences, conventions) that can also influence the interpretation of a visual representation~\cite{Hullman2011} and, thus, the decisions of the viewer.

Our findings may also be limited by the background of our participants. CS students are familiarized with visual representations of data and scientific concepts. Additional studies are needed to explore whether the effects we observed hold for students from different backgrounds.

Moreover, our observations of the grades maximization effect are not conclusive. Given the evidence we gathered, our intuition is that, in the presence of predictions for the GPA, students tend to look for maximization, but not when only individual course grades are shown. This question, however, is yet to be answered. 
Additional studies (e.g., in-situ pilots) are also needed to issue more concrete design recommendations for future course selection tools. \app~is a high-fidelity prototype, but its institutional deployment is still subject to the outcomes of multiple studies and discussions with several parties. The use of learning analytics dashboards should be thoroughly analyzed before their adoption at HEIs. 

Finally, our findings highlight that the design of a tool like \app~must consider the role that both students and advisors play in the course selection process. In line with the goal of interactive visualization technologies, \app~and similar tools require humans in the loop and this requirement should not be underestimated. Otherwise, this type of technologies run the risk of being perceived as oracles that people are supposed to trust and never question.






\section{Conclusion}

This paper investigated the effects of performance predictions on students when they plan their upcoming term. To this end, we used \app, an interactive visualization tool that enables the composition of arbitrary sets of courses and provides performance predictions and explanations.

A qualitative study of the tool found that in response to performance predictions for both individual course grades and the GPA, students tend to approach course selection as a performance maximization problem, even to the detriment of other factors such as the workload. We also observed little interest in understanding the rationale behind the predictions provided by the tool.

In a follow-up quantitative study, we investigated whether the maximization and overreliance effects were affected by the type of visual representation used to convey \app's performance predictions. To this end, we designed a \textit{specific} to \textit{vague} spectrum of visual representations for performance predictions. In this second study, we did not found evidence of maximization effects when the GPA is not shown together with the individual course grade predictions. The participants' lack of interest in the explanations, however, persisted. We also found several significant differences in aspects such as the average predicted grade and workload of the selected courses. These differences arose both among visual representations of the same type and between different types.

Our observations show that framing effects arise when visual structures are used to communicate performance predictions to students. That is, some of the visual representations we studied have the potential to shape the students' decisions and their decision process. Furthermore, specific types of visual representations elicit strong preferences and aversions on the students. These observations are of great value to design better data-driven course selection tools. Equally importantly, our insights provide new empirical evidence on how different design choices can shape the way people interpret visual representations of data.

\begin{acks}

The authors wish to thank the reviewers and associate chairs for their advice and insightful comments. We are also indebted to the students who donated their time to participate in our studies.

This work was funded by the LALA project (grant no. 586120-EPP-1-2017-1-ES-EPPKA2-CBHE-JP), and with support from the European Commission. This publication reflects only the views of the authors, and the Commission cannot be held responsible for any use which may be made of the information contained therein.

\end{acks}

\balance{}

\bibliographystyle{ACM-Reference-Format}
\bibliography{bibliography}


\begin{thebibliography}{70}


\ifx \showCODEN    \undefined \def \showCODEN     #1{\unskip}     \fi
\ifx \showDOI      \undefined \def \showDOI       #1{#1}\fi
\ifx \showISBNx    \undefined \def \showISBNx     #1{\unskip}     \fi
\ifx \showISBNxiii \undefined \def \showISBNxiii  #1{\unskip}     \fi
\ifx \showISSN     \undefined \def \showISSN      #1{\unskip}     \fi
\ifx \showLCCN     \undefined \def \showLCCN      #1{\unskip}     \fi
\ifx \shownote     \undefined \def \shownote      #1{#1}          \fi
\ifx \showarticletitle \undefined \def \showarticletitle #1{#1}   \fi
\ifx \showURL      \undefined \def \showURL       {\relax}        \fi
\providecommand\bibfield[2]{#2}
\providecommand\bibinfo[2]{#2}
\providecommand\natexlab[1]{#1}
\providecommand\showeprint[2][]{arXiv:#2}

\bibitem[\protect\citeauthoryear{Agency}{Agency}{2001}]%
        {QAACode}
\bibfield{author}{\bibinfo{person}{The Quality~Assurance Agency}.}
  \bibinfo{year}{2001}\natexlab{}.
\newblock \bibinfo{booktitle}{\emph{QAA Code of Practice for the assurance of
  academic quality and standards in higher education: Career Education,
  Information and Guidance (CEIG)}}.
\newblock Taylor \& Francis.
\newblock
\urldef\tempurl%
\url{https://doi.org/10.11120/plan.2001.00020026}
\showDOI{\tempurl}


\bibitem[\protect\citeauthoryear{Aguilar, Karabenick, Teasley, and
  Baek}{Aguilar et~al\mbox{.}}{2021}]%
        {Aguilar2020associations}
\bibfield{author}{\bibinfo{person}{Stephen~J. Aguilar},
  \bibinfo{person}{Stuart~A. Karabenick}, \bibinfo{person}{Stephanie~D.
  Teasley}, {and} \bibinfo{person}{Clare Baek}.}
  \bibinfo{year}{2021}\natexlab{}.
\newblock \showarticletitle{{Associations between Learning Analytics Dashboard
  Exposure and Motivation and Self-regulated Learning}}.
\newblock \bibinfo{journal}{\emph{Computers \& Education}}
  \bibinfo{volume}{162} (\bibinfo{year}{2021}), \bibinfo{pages}{104085}.
\newblock
\showISSN{0360-1315}
\urldef\tempurl%
\url{https://doi.org/10.1016/j.compedu.2020.104085}
\showDOI{\tempurl}


\bibitem[\protect\citeauthoryear{Atapattu, Falkner, and Tarmazdi}{Atapattu
  et~al\mbox{.}}{2016}]%
        {atapattu2016topic}
\bibfield{author}{\bibinfo{person}{Thushari Atapattu}, \bibinfo{person}{Katrina
  Falkner}, {and} \bibinfo{person}{Hamid Tarmazdi}.}
  \bibinfo{year}{2016}\natexlab{}.
\newblock \showarticletitle{{Topic-Wise Classification of MOOC Discussions: A
  Visual Analytics Approach}}. In \bibinfo{booktitle}{\emph{Proceedings of the
  9th International Conference on Educational Data Mining}}.
  \bibinfo{publisher}{International Educational Data Mining Society}.
\newblock


\bibitem[\protect\citeauthoryear{Bandura, Rumsey, Walker, and Harris}{Bandura
  et~al\mbox{.}}{1994}]%
        {Bandura1994}
\bibfield{author}{\bibinfo{person}{Albert Bandura}, \bibinfo{person}{M Rumsey},
  \bibinfo{person}{C Walker}, {and} \bibinfo{person}{J Harris}.}
  \bibinfo{year}{1994}\natexlab{}.
\newblock \showarticletitle{Regulative function of perceived self-efficacy}.
\newblock \bibinfo{journal}{\emph{Personnel selection and classification}}
  (\bibinfo{year}{1994}), \bibinfo{pages}{261--271}.
\newblock


\bibitem[\protect\citeauthoryear{Bar, Kadiyali, and Zussman}{Bar
  et~al\mbox{.}}{2008}]%
        {Bar2005QuestFK}
\bibfield{author}{\bibinfo{person}{Talia Bar}, \bibinfo{person}{Vrinda
  Kadiyali}, {and} \bibinfo{person}{Asaf Zussman}.}
  \bibinfo{year}{2008}\natexlab{}.
\newblock \showarticletitle{Quest for Knowledge and Pursuit of Grades:
  Information, Course Selection, and Grade Inflation}.
\newblock \bibinfo{journal}{\emph{Behavioral \& Experimental Economics}}
  (\bibinfo{year}{2008}).
\newblock
\urldef\tempurl%
\url{https://doi.org/10.2139/ssrn.1019580}
\showDOI{\tempurl}


\bibitem[\protect\citeauthoryear{Bateman, Mandryk, Gutwin, Genest, McDine, and
  Brooks}{Bateman et~al\mbox{.}}{2010}]%
        {Bateman}
\bibfield{author}{\bibinfo{person}{Scott Bateman}, \bibinfo{person}{Regan~L.
  Mandryk}, \bibinfo{person}{Carl Gutwin}, \bibinfo{person}{Aaron Genest},
  \bibinfo{person}{David McDine}, {and} \bibinfo{person}{Christopher Brooks}.}
  \bibinfo{year}{2010}\natexlab{}.
\newblock \showarticletitle{Useful {Junk}?: {The} {Effects} of {Visual}
  {Embellishment} on {Comprehension} and {Memorability} of {Charts}}. In
  \bibinfo{booktitle}{\emph{Proceedings of the {SIGCHI} {Conference} on {Human}
  {Factors} in {Computing} {Systems}}} \emph{(\bibinfo{series}{{CHI} '10})}.
  \bibinfo{publisher}{ACM}, \bibinfo{address}{New York, NY, USA},
  \bibinfo{pages}{2573--2582}.
\newblock
\showISBNx{978-1-60558-929-9}
\urldef\tempurl%
\url{https://doi.org/10.1145/1753326.1753716}
\showDOI{\tempurl}


\bibitem[\protect\citeauthoryear{Baumer, Polletta, Pierski, and Gay}{Baumer
  et~al\mbox{.}}{2017}]%
        {baumer2017simple}
\bibfield{author}{\bibinfo{person}{Eric P.~S. Baumer},
  \bibinfo{person}{Francesca Polletta}, \bibinfo{person}{Nicole Pierski}, {and}
  \bibinfo{person}{Geri~K. Gay}.} \bibinfo{year}{2017}\natexlab{}.
\newblock \showarticletitle{A simple intervention to reduce framing effects in
  perceptions of global climate change}.
\newblock \bibinfo{journal}{\emph{Environmental Communication}}
  \bibinfo{volume}{11}, \bibinfo{number}{3} (\bibinfo{year}{2017}),
  \bibinfo{pages}{289--310}.
\newblock
\urldef\tempurl%
\url{https://doi.org/10.1080/17524032.2015.1084015}
\showDOI{\tempurl}


\bibitem[\protect\citeauthoryear{Baumer, Snyder, and Gay}{Baumer
  et~al\mbox{.}}{2018}]%
        {Baumer2018}
\bibfield{author}{\bibinfo{person}{Eric P.~S. Baumer}, \bibinfo{person}{Jaime
  Snyder}, {and} \bibinfo{person}{Geri~K. Gay}.}
  \bibinfo{year}{2018}\natexlab{}.
\newblock \showarticletitle{Interpretive Impacts of Text Visualization:
  Mitigating Political Framing Effects}.
\newblock \bibinfo{journal}{\emph{ACM Transactions on Computer-Human
  Interaction}} \bibinfo{volume}{25}, \bibinfo{number}{4}, Article
  \bibinfo{articleno}{20} (\bibinfo{date}{Aug.} \bibinfo{year}{2018}),
  \bibinfo{numpages}{26}~pages.
\newblock
\showISSN{1073-0516}
\urldef\tempurl%
\url{https://doi.org/10.1145/3214353}
\showDOI{\tempurl}


\bibitem[\protect\citeauthoryear{Beattie, Woodley, and Souter}{Beattie
  et~al\mbox{.}}{2014}]%
        {beattie2014creepy}
\bibfield{author}{\bibinfo{person}{Scott Beattie}, \bibinfo{person}{Carolyn
  Woodley}, {and} \bibinfo{person}{Kay Souter}.}
  \bibinfo{year}{2014}\natexlab{}.
\newblock \showarticletitle{Creepy analytics and learner data rights}.
\newblock \bibinfo{journal}{\emph{Rhetoric and reality: Critical perspectives
  on educational technology. Proceedings ascilite}} (\bibinfo{year}{2014}),
  \bibinfo{pages}{421--425}.
\newblock
\urldef\tempurl%
\url{http://www.academia.edu/download/35842981/69-Beattie.pdf}
\showURL{%
\tempurl}


\bibitem[\protect\citeauthoryear{Beheshitha, Hatala, Ga\v{s}evi\'{c}, and
  Joksimovi\'{c}}{Beheshitha et~al\mbox{.}}{2016}]%
        {10.1145/2883851.2883904}
\bibfield{author}{\bibinfo{person}{Sanam~Shirazi Beheshitha},
  \bibinfo{person}{Marek Hatala}, \bibinfo{person}{Dragan Ga\v{s}evi\'{c}},
  {and} \bibinfo{person}{Sre\'{c}ko Joksimovi\'{c}}.}
  \bibinfo{year}{2016}\natexlab{}.
\newblock \showarticletitle{The Role of Achievement Goal Orientations When
  Studying Effect of Learning Analytics Visualizations}. In
  \bibinfo{booktitle}{\emph{Proceedings of the Sixth International Conference
  on Learning Analytics \& Knowledge}} (Edinburgh, United Kingdom)
  \emph{(\bibinfo{series}{LAK '16})}. \bibinfo{publisher}{Association for
  Computing Machinery}, \bibinfo{address}{New York, NY, USA},
  \bibinfo{pages}{54–63}.
\newblock
\showISBNx{9781450341905}
\urldef\tempurl%
\url{https://doi.org/10.1145/2883851.2883904}
\showDOI{\tempurl}


\bibitem[\protect\citeauthoryear{Bodily and Verbert}{Bodily and
  Verbert}{2017}]%
        {8010828}
\bibfield{author}{\bibinfo{person}{Robert Bodily} {and}
  \bibinfo{person}{Katrien Verbert}.} \bibinfo{year}{2017}\natexlab{}.
\newblock \showarticletitle{Review of Research on Student-Facing Learning
  Analytics Dashboards and Educational Recommender Systems}.
\newblock \bibinfo{journal}{\emph{IEEE Transactions on Learning Technologies}}
  \bibinfo{volume}{10}, \bibinfo{number}{4} (\bibinfo{year}{2017}),
  \bibinfo{pages}{405--418}.
\newblock
\urldef\tempurl%
\url{https://doi.org/10.1109/TLT.2017.2740172}
\showDOI{\tempurl}


\bibitem[\protect\citeauthoryear{Boy, Pandey, Emerson, Satterthwaite, Nov, and
  Bertini}{Boy et~al\mbox{.}}{2017}]%
        {Boy2017PeopleBehindData}
\bibfield{author}{\bibinfo{person}{Jeremy Boy}, \bibinfo{person}{Anshul~Vikram
  Pandey}, \bibinfo{person}{John Emerson}, \bibinfo{person}{Margaret
  Satterthwaite}, \bibinfo{person}{Oded Nov}, {and} \bibinfo{person}{Enrico
  Bertini}.} \bibinfo{year}{2017}\natexlab{}.
\newblock \showarticletitle{Showing People Behind Data: Does Anthropomorphizing
  Visualizations Elicit More Empathy for Human Rights Data?}. In
  \bibinfo{booktitle}{\emph{Proceedings of the 2017 CHI Conference on Human
  Factors in Computing Systems}} (Denver, Colorado, USA)
  \emph{(\bibinfo{series}{CHI '17})}. \bibinfo{publisher}{Association for
  Computing Machinery}, \bibinfo{address}{New York, NY, USA},
  \bibinfo{pages}{5462–5474}.
\newblock
\showISBNx{9781450346559}
\urldef\tempurl%
\url{https://doi.org/10.1145/3025453.3025512}
\showDOI{\tempurl}


\bibitem[\protect\citeauthoryear{Boyatzis}{Boyatzis}{1998}]%
        {boyatzis1998}
\bibfield{author}{\bibinfo{person}{Richard~E Boyatzis}.}
  \bibinfo{year}{1998}\natexlab{}.
\newblock \bibinfo{booktitle}{\emph{{Transforming Qualitative Information:
  Thematic Analysis and Code Development}}}.
\newblock \bibinfo{publisher}{Sage}.
\newblock


\bibitem[\protect\citeauthoryear{Bull, Kickmeier-Rust, Vatrapu, Johnson,
  Hammermueller, Byrne, Hernandez-Munoz, Giorgini, and Meissl-Egghart}{Bull
  et~al\mbox{.}}{2013}]%
        {10.1007/978-3-642-40814-4_51}
\bibfield{author}{\bibinfo{person}{Susan Bull}, \bibinfo{person}{Michael
  Kickmeier-Rust}, \bibinfo{person}{Ravi~K. Vatrapu},
  \bibinfo{person}{Matthew~D. Johnson}, \bibinfo{person}{Klaus Hammermueller},
  \bibinfo{person}{William Byrne}, \bibinfo{person}{Luis Hernandez-Munoz},
  \bibinfo{person}{Fabrizio Giorgini}, {and} \bibinfo{person}{Gerhilde
  Meissl-Egghart}.} \bibinfo{year}{2013}\natexlab{}.
\newblock \showarticletitle{Learning, Learning Analytics, Activity
  Visualisation and Open Learner Model: Confusing?}. In
  \bibinfo{booktitle}{\emph{Scaling up Learning for Sustained Impact}},
  \bibfield{editor}{\bibinfo{person}{Davinia Hern{\'a}ndez-Leo},
  \bibinfo{person}{Tobias Ley}, \bibinfo{person}{Ralf Klamma}, {and}
  \bibinfo{person}{Andreas Harrer}} (Eds.). \bibinfo{publisher}{Springer Berlin
  Heidelberg}, \bibinfo{address}{Berlin, Heidelberg},
  \bibinfo{pages}{532--535}.
\newblock
\showISBNx{978-3-642-40814-4}
\urldef\tempurl%
\url{https://doi.org/10.1007/978-3-642-40814-4_51}
\showDOI{\tempurl}


\bibitem[\protect\citeauthoryear{Castells, Doust, Gal{\'a}rraga, M{\'e}ndez,
  Ortiz-Rojas, and Jim{\'e}nez}{Castells et~al\mbox{.}}{2020}]%
        {Castells2020}
\bibfield{author}{\bibinfo{person}{Jaime Castells},
  \bibinfo{person}{Mohammad~Poul Doust}, \bibinfo{person}{Luis Gal{\'a}rraga},
  \bibinfo{person}{Gonzalo~Gabriel M{\'e}ndez}, \bibinfo{person}{Margarita
  Ortiz-Rojas}, {and} \bibinfo{person}{Alberto Jim{\'e}nez}.}
  \bibinfo{year}{2020}\natexlab{}.
\newblock \showarticletitle{A Student-oriented Tool to Support Course Selection
  in Academic Counseling Sessions}. In \bibinfo{booktitle}{\emph{Proceedings of
  the Workshop on Adoption, Adaptation and Pilots of Learning Analytics in
  Under-represented Regions, co-located with the 15th European Conference on
  Technology Enhanced Learning 2020 (ECTEL 2020)}}.
\newblock
\urldef\tempurl%
\url{http://ceur-ws.org/Vol-2704/paper4.pdf}
\showURL{%
\tempurl}


\bibitem[\protect\citeauthoryear{Caulkins, Larkey, and Wei}{Caulkins
  et~al\mbox{.}}{1996}]%
        {Caulkins:1996:AdjustingGPA}
\bibfield{author}{\bibinfo{person}{Jonathan~P Caulkins},
  \bibinfo{person}{Patrick~D. Larkey}, {and} \bibinfo{person}{Jifa Wei}.}
  \bibinfo{year}{1996}\natexlab{}.
\newblock \showarticletitle{{Adjusting GPA to Reflect Course Difficulty}}.
\newblock  (\bibinfo{date}{1} \bibinfo{year}{1996}).
\newblock
\urldef\tempurl%
\url{https://doi.org/10.1184/R1/6470981.v1}
\showDOI{\tempurl}


\bibitem[\protect\citeauthoryear{Chamorro-Premuzic and
  Furnham}{Chamorro-Premuzic and Furnham}{2008}]%
        {Chamorro2008}
\bibfield{author}{\bibinfo{person}{Tomas Chamorro-Premuzic} {and}
  \bibinfo{person}{Adrian Furnham}.} \bibinfo{year}{2008}\natexlab{}.
\newblock \showarticletitle{Personality, intelligence and approaches to
  learning as predictors of academic performance}.
\newblock \bibinfo{journal}{\emph{Personality and Individual Differences}}
  \bibinfo{volume}{44}, \bibinfo{number}{7} (\bibinfo{year}{2008}),
  \bibinfo{pages}{1596 -- 1603}.
\newblock
\showISSN{0191-8869}
\urldef\tempurl%
\url{https://doi.org/10.1016/j.paid.2008.01.003}
\showDOI{\tempurl}


\bibitem[\protect\citeauthoryear{Charleer, Moere, Klerkx, Verbert, and
  Laet}{Charleer et~al\mbox{.}}{2018}]%
        {Charleer:2018:LearningAnalyticsDashboards}
\bibfield{author}{\bibinfo{person}{Sven Charleer},
  \bibinfo{person}{Andrew~Vande Moere}, \bibinfo{person}{Joris Klerkx},
  \bibinfo{person}{Katrien Verbert}, {and} \bibinfo{person}{Tinne~De Laet}.}
  \bibinfo{year}{2018}\natexlab{}.
\newblock \showarticletitle{{Learning Analytics Dashboards to Support
  Adviser-Student Dialogue}}.
\newblock \bibinfo{journal}{\emph{IEEE Transactions on Learning Technologies}}
  \bibinfo{volume}{11}, \bibinfo{number}{3} (\bibinfo{year}{2018}),
  \bibinfo{pages}{389--399}.
\newblock
\urldef\tempurl%
\url{https://doi.org/10.1109/TLT.2017.2720670}
\showDOI{\tempurl}


\bibitem[\protect\citeauthoryear{Chaturapruek, Dee, Johari, Kizilcec, and
  Stevens}{Chaturapruek et~al\mbox{.}}{2018}]%
        {Chaturapruek2018}
\bibfield{author}{\bibinfo{person}{Sorathan Chaturapruek},
  \bibinfo{person}{Thomas~S. Dee}, \bibinfo{person}{Ramesh Johari},
  \bibinfo{person}{Ren\'{e}~F. Kizilcec}, {and} \bibinfo{person}{Mitchell~L.
  Stevens}.} \bibinfo{year}{2018}\natexlab{}.
\newblock \showarticletitle{How a Data-Driven Course Planning Tool Affects
  College Students’ GPA: Evidence from Two Field Experiments}. In
  \bibinfo{booktitle}{\emph{Proceedings of the Fifth Annual ACM Conference on
  Learning at Scale}} (London, United Kingdom) \emph{(\bibinfo{series}{L@S
  ’18})}. \bibinfo{publisher}{Association for Computing Machinery},
  \bibinfo{address}{New York, NY, USA}, Article \bibinfo{articleno}{63},
  \bibinfo{numpages}{10}~pages.
\newblock
\showISBNx{9781450358866}
\urldef\tempurl%
\url{https://doi.org/10.1145/3231644.3231668}
\showDOI{\tempurl}


\bibitem[\protect\citeauthoryear{Cheema and Bagchi}{Cheema and Bagchi}{2011}]%
        {Cheema2011EffectGoalVisualizationOnGoalPursuit}
\bibfield{author}{\bibinfo{person}{Amar Cheema} {and} \bibinfo{person}{Rajesh
  Bagchi}.} \bibinfo{year}{2011}\natexlab{}.
\newblock \showarticletitle{The Effect of Goal Visualization on Goal Pursuit:
  Implications for Consumers and Managers}.
\newblock \bibinfo{journal}{\emph{Journal of Marketing}} \bibinfo{volume}{75},
  \bibinfo{number}{2} (\bibinfo{year}{2011}), \bibinfo{pages}{109--123}.
\newblock
\urldef\tempurl%
\url{https://doi.org/10.1509/jm.75.2.109}
\showDOI{\tempurl}
\showeprint{https://doi.org/10.1509/jm.75.2.109}


\bibitem[\protect\citeauthoryear{Chernoff}{Chernoff}{1973}]%
        {Chernoff1973}
\bibfield{author}{\bibinfo{person}{Herman Chernoff}.}
  \bibinfo{year}{1973}\natexlab{}.
\newblock \showarticletitle{The Use of Faces to Represent Points in
  k-Dimensional Space Graphically}.
\newblock \bibinfo{journal}{\emph{J. Amer. Statist. Assoc.}}
  \bibinfo{volume}{68}, \bibinfo{number}{342} (\bibinfo{year}{1973}),
  \bibinfo{pages}{361--368}.
\newblock
\urldef\tempurl%
\url{https://doi.org/10.1080/01621459.1973.10482434}
\showDOI{\tempurl}


\bibitem[\protect\citeauthoryear{Chong and Druckman}{Chong and
  Druckman}{2007}]%
        {chong2007theory}
\bibfield{author}{\bibinfo{person}{Dennis Chong} {and} \bibinfo{person}{James~N
  Druckman}.} \bibinfo{year}{2007}\natexlab{}.
\newblock \showarticletitle{A theory of framing and opinion formation in
  competitive elite environments}.
\newblock \bibinfo{journal}{\emph{Journal of communication}}
  \bibinfo{volume}{57}, \bibinfo{number}{1} (\bibinfo{year}{2007}),
  \bibinfo{pages}{99--118}.
\newblock
\urldef\tempurl%
\url{https://doi.org/10.1111/j.1460-2466.2006.00331.x}
\showDOI{\tempurl}


\bibitem[\protect\citeauthoryear{Cleveland and McGill}{Cleveland and
  McGill}{1984}]%
        {Cleveland1984}
\bibfield{author}{\bibinfo{person}{William~S Cleveland} {and}
  \bibinfo{person}{Robert McGill}.} \bibinfo{year}{1984}\natexlab{}.
\newblock \showarticletitle{{Graphical Perception: Theory, Experimentation, and
  Application to the Development of Graphical Methods}}.
\newblock \bibinfo{journal}{\emph{J. Amer. Statist. Assoc.}}
  \bibinfo{volume}{79}, \bibinfo{number}{387} (\bibinfo{year}{1984}),
  \bibinfo{pages}{531--554}.
\newblock
\showISBNx{0162-1459}
\showISSN{01621459}
\urldef\tempurl%
\url{https://doi.org/10.2307/2288400}
\showDOI{\tempurl}
\showeprint[arxiv]{arXiv:1011.1669v3}


\bibitem[\protect\citeauthoryear{Cleveland and McGill}{Cleveland and
  McGill}{1987}]%
        {Cleveland1987}
\bibfield{author}{\bibinfo{person}{William~S. Cleveland} {and}
  \bibinfo{person}{Robert McGill}.} \bibinfo{year}{1987}\natexlab{}.
\newblock \showarticletitle{Graphical Perception: The Visual Decoding of
  Quantitative Information on Graphical Displays of Data}.
\newblock \bibinfo{journal}{\emph{Journal of the Royal Statistical Society.
  Series A (General)}} \bibinfo{volume}{150}, \bibinfo{number}{3}
  (\bibinfo{year}{1987}), \bibinfo{pages}{pp. 192--229}.
\newblock
\showISSN{00359238}
\urldef\tempurl%
\url{http://www.jstor.org/stable/2981473}
\showURL{%
\tempurl}


\bibitem[\protect\citeauthoryear{Druckman}{Druckman}{2001}]%
        {druckman2001implications}
\bibfield{author}{\bibinfo{person}{James~N Druckman}.}
  \bibinfo{year}{2001}\natexlab{}.
\newblock \showarticletitle{The implications of framing effects for citizen
  competence}.
\newblock \bibinfo{journal}{\emph{Political behavior}} \bibinfo{volume}{23},
  \bibinfo{number}{3} (\bibinfo{year}{2001}), \bibinfo{pages}{225--256}.
\newblock
\urldef\tempurl%
\url{https://doi.org/10.1023/A:1015006907312}
\showDOI{\tempurl}


\bibitem[\protect\citeauthoryear{Fleur, van~den Bos, and Bredeweg}{Fleur
  et~al\mbox{.}}{2020}]%
        {fleur2020learning}
\bibfield{author}{\bibinfo{person}{Damien~S Fleur}, \bibinfo{person}{Wouter
  van~den Bos}, {and} \bibinfo{person}{Bert Bredeweg}.}
  \bibinfo{year}{2020}\natexlab{}.
\newblock \showarticletitle{Learning Analytics Dashboard for Motivation and
  Performance}. In \bibinfo{booktitle}{\emph{International Conference on
  Intelligent Tutoring Systems}}. Springer, \bibinfo{pages}{411--419}.
\newblock


\bibitem[\protect\citeauthoryear{Fürnkranz, Kliegr, and Paulheim}{Fürnkranz
  et~al\mbox{.}}{2019}]%
        {Frnkranz2019OnCP}
\bibfield{author}{\bibinfo{person}{Johannes Fürnkranz},
  \bibinfo{person}{Tomáš Kliegr}, {and} \bibinfo{person}{Heiko Paulheim}.}
  \bibinfo{year}{2019}\natexlab{}.
\newblock \showarticletitle{On cognitive preferences and the plausibility of
  rule-based models}.
\newblock \bibinfo{journal}{\emph{Machine Learning}}  \bibinfo{volume}{109}
  (\bibinfo{year}{2019}), \bibinfo{pages}{853--898}.
\newblock
\urldef\tempurl%
\url{https://doi.org/10.1007/s10994-019-05856-5}
\showDOI{\tempurl}


\bibitem[\protect\citeauthoryear{Ga\v{s}evi\'{c}, Zouaq, and
  Janzen}{Ga\v{s}evi\'{c} et~al\mbox{.}}{2013}]%
        {Gasevic2013}
\bibfield{author}{\bibinfo{person}{Dragan Ga\v{s}evi\'{c}},
  \bibinfo{person}{Amal Zouaq}, {and} \bibinfo{person}{Robert Janzen}.}
  \bibinfo{year}{2013}\natexlab{}.
\newblock \showarticletitle{“Choose Your Classmates, Your GPA Is at
  Stake!”: The Association of Cross-Class Social Ties and Academic
  Performance}.
\newblock \bibinfo{journal}{\emph{American Behavioral Scientist}}
  \bibinfo{volume}{57}, \bibinfo{number}{10} (\bibinfo{year}{2013}),
  \bibinfo{pages}{1460--1479}.
\newblock
\urldef\tempurl%
\url{https://doi.org/10.1177/0002764213479362}
\showDOI{\tempurl}


\bibitem[\protect\citeauthoryear{Goddard, Roudsari, and Wyatt}{Goddard
  et~al\mbox{.}}{2012}]%
        {goddard2012automation}
\bibfield{author}{\bibinfo{person}{Kate Goddard}, \bibinfo{person}{Abdul
  Roudsari}, {and} \bibinfo{person}{Jeremy~C Wyatt}.}
  \bibinfo{year}{2012}\natexlab{}.
\newblock \showarticletitle{Automation bias: a systematic review of frequency,
  effect mediators, and mitigators}.
\newblock \bibinfo{journal}{\emph{Journal of the American Medical Informatics
  Association}} \bibinfo{volume}{19}, \bibinfo{number}{1}
  (\bibinfo{year}{2012}), \bibinfo{pages}{121--127}.
\newblock
\urldef\tempurl%
\url{https://doi.org/10.1136/amiajnl-2011-000089}
\showDOI{\tempurl}


\bibitem[\protect\citeauthoryear{Green and Celkan}{Green and Celkan}{2011}]%
        {Green2011StudentDemographic}
\bibfield{author}{\bibinfo{person}{Linda Green} {and} \bibinfo{person}{Gul
  Celkan}.} \bibinfo{year}{2011}\natexlab{}.
\newblock \showarticletitle{Student demographic characteristics and how they
  relate to student achievement}.
\newblock \bibinfo{journal}{\emph{Procedia - Social and Behavioral Sciences}}
  \bibinfo{volume}{15} (\bibinfo{year}{2011}), \bibinfo{pages}{341 -- 345}.
\newblock
\showISSN{1877-0428}
\urldef\tempurl%
\url{https://doi.org/10.1016/j.sbspro.2011.03.098}
\showDOI{\tempurl}


\bibitem[\protect\citeauthoryear{Greller and Drachsler}{Greller and
  Drachsler}{2012}]%
        {greller2012translating}
\bibfield{author}{\bibinfo{person}{Wolfgang Greller} {and}
  \bibinfo{person}{Hendrik Drachsler}.} \bibinfo{year}{2012}\natexlab{}.
\newblock \showarticletitle{Translating learning into numbers: A generic
  framework for learning analytics}.
\newblock \bibinfo{journal}{\emph{Journal of Educational Technology \&
  Society}} \bibinfo{volume}{15}, \bibinfo{number}{3} (\bibinfo{year}{2012}),
  \bibinfo{pages}{42--57}.
\newblock


\bibitem[\protect\citeauthoryear{Grupe}{Grupe}{2002}]%
        {grupe2002internet}
\bibfield{author}{\bibinfo{person}{Fritz~H Grupe}.}
  \bibinfo{year}{2002}\natexlab{}.
\newblock \showarticletitle{An Internet-based expert system for selecting an
  academic major: www.MyMajors.com}.
\newblock \bibinfo{journal}{\emph{The Internet and Higher Education}}
  \bibinfo{volume}{5}, \bibinfo{number}{4} (\bibinfo{year}{2002}),
  \bibinfo{pages}{333 -- 344}.
\newblock
\showISSN{1096-7516}
\urldef\tempurl%
\url{https://doi.org/10.1016/S1096-7516(02)00129-X}
\showDOI{\tempurl}


\bibitem[\protect\citeauthoryear{Gutiérrez, Seipp, Ochoa, Chiluiza, Laet, and
  Verbert}{Gutiérrez et~al\mbox{.}}{2018}]%
        {guttierez2018LADA}
\bibfield{author}{\bibinfo{person}{Francisco Gutiérrez},
  \bibinfo{person}{Karsten Seipp}, \bibinfo{person}{Xavier Ochoa},
  \bibinfo{person}{Katherine Chiluiza}, \bibinfo{person}{Tinne~De Laet}, {and}
  \bibinfo{person}{Katrien Verbert}.} \bibinfo{year}{2018}\natexlab{}.
\newblock \showarticletitle{{LADA: A Learning Analytics Dashboard for Academic
  Advising}}.
\newblock \bibinfo{journal}{\emph{Computers in Human Behavior}}
  \bibinfo{volume}{107} (\bibinfo{year}{2018}).
\newblock
\urldef\tempurl%
\url{https://doi.org/10.1016/j.chb.2018.12.004}
\showDOI{\tempurl}


\bibitem[\protect\citeauthoryear{Hullman, Adar, and Shah}{Hullman
  et~al\mbox{.}}{2011}]%
        {Hullman2011}
\bibfield{author}{\bibinfo{person}{Jessica Hullman}, \bibinfo{person}{Eytan
  Adar}, {and} \bibinfo{person}{Priti Shah}.} \bibinfo{year}{2011}\natexlab{}.
\newblock \showarticletitle{Benefitting InfoVis with Visual Difficulties}.
\newblock \bibinfo{journal}{\emph{IEEE Transactions on Visualization and
  Computer Graphics}} \bibinfo{volume}{17}, \bibinfo{number}{12}
  (\bibinfo{date}{Dec} \bibinfo{year}{2011}), \bibinfo{pages}{2213--2222}.
\newblock
\showISSN{1077-2626}
\urldef\tempurl%
\url{https://doi.org/10.1109/TVCG.2011.175}
\showDOI{\tempurl}


\bibitem[\protect\citeauthoryear{Jonassen}{Jonassen}{1994}]%
        {Jonassen1994Thinking}
\bibfield{author}{\bibinfo{person}{David~H. Jonassen}.}
  \bibinfo{year}{1994}\natexlab{}.
\newblock \showarticletitle{Thinking Technology: Toward a Constructivist Design
  Model}.
\newblock \bibinfo{journal}{\emph{Educational Technology}}
  \bibinfo{volume}{34}, \bibinfo{number}{4} (\bibinfo{year}{1994}),
  \bibinfo{pages}{34--37}.
\newblock
\showISSN{0013-1962}
\urldef\tempurl%
\url{https://www.learntechlib.org/p/171050}
\showURL{%
\tempurl}


\bibitem[\protect\citeauthoryear{Kardan, Sadeghi, Ghidary, and Sani}{Kardan
  et~al\mbox{.}}{2013}]%
        {KARDAN20131}
\bibfield{author}{\bibinfo{person}{Ahmad~A. Kardan}, \bibinfo{person}{Hamid
  Sadeghi}, \bibinfo{person}{Saeed~Shiry Ghidary}, {and}
  \bibinfo{person}{Mohammad Reza~Fani Sani}.} \bibinfo{year}{2013}\natexlab{}.
\newblock \showarticletitle{Prediction of student course selection in online
  higher education institutes using neural network}.
\newblock \bibinfo{journal}{\emph{Computers and Education}}
  \bibinfo{volume}{65} (\bibinfo{year}{2013}), \bibinfo{pages}{1 -- 11}.
\newblock
\showISSN{0360-1315}
\urldef\tempurl%
\url{https://doi.org/10.1016/j.compedu.2013.01.015}
\showDOI{\tempurl}


\bibitem[\protect\citeauthoryear{Klomegah}{Klomegah}{2007}]%
        {klomegah2007predictors}
\bibfield{author}{\bibinfo{person}{Roger~Yao Klomegah}.}
  \bibinfo{year}{2007}\natexlab{}.
\newblock \showarticletitle{Predictors of Academic Performance of University
  Students: An Application of the Goal Efficacy Model.}
\newblock \bibinfo{journal}{\emph{College Student Journal}}
  \bibinfo{volume}{41}, \bibinfo{number}{2} (\bibinfo{year}{2007}).
\newblock


\bibitem[\protect\citeauthoryear{Kong, Liu, and Karahalios}{Kong
  et~al\mbox{.}}{2018}]%
        {Kong2018FramesAndSlants}
\bibfield{author}{\bibinfo{person}{Ha-Kyung Kong}, \bibinfo{person}{Zhicheng
  Liu}, {and} \bibinfo{person}{Karrie Karahalios}.}
  \bibinfo{year}{2018}\natexlab{}.
\newblock \showarticletitle{Frames and Slants in Titles of Visualizations on
  Controversial Topics}. In \bibinfo{booktitle}{\emph{Proceedings of the 2018
  CHI Conference on Human Factors in Computing Systems}} (Montreal QC, Canada)
  \emph{(\bibinfo{series}{CHI '18})}. \bibinfo{publisher}{Association for
  Computing Machinery}, \bibinfo{address}{New York, NY, USA},
  \bibinfo{pages}{1–12}.
\newblock
\showISBNx{9781450356206}
\urldef\tempurl%
\url{https://doi.org/10.1145/3173574.3174012}
\showDOI{\tempurl}


\bibitem[\protect\citeauthoryear{Kong, Liu, and Karahalios}{Kong
  et~al\mbox{.}}{2019}]%
        {Kong2019TrustAndRecall}
\bibfield{author}{\bibinfo{person}{Ha-Kyung Kong}, \bibinfo{person}{Zhicheng
  Liu}, {and} \bibinfo{person}{Karrie Karahalios}.}
  \bibinfo{year}{2019}\natexlab{}.
\newblock \showarticletitle{Trust and Recall of Information across Varying
  Degrees of Title-Visualization Misalignment}. In
  \bibinfo{booktitle}{\emph{Proceedings of the 2019 CHI Conference on Human
  Factors in Computing Systems}} (Glasgow, Scotland Uk)
  \emph{(\bibinfo{series}{CHI '19})}. \bibinfo{publisher}{Association for
  Computing Machinery}, \bibinfo{address}{New York, NY, USA},
  \bibinfo{pages}{1–13}.
\newblock
\showISBNx{9781450359702}
\urldef\tempurl%
\url{https://doi.org/10.1145/3290605.3300576}
\showDOI{\tempurl}


\bibitem[\protect\citeauthoryear{Levin, Schneider, and Gaeth}{Levin
  et~al\mbox{.}}{1998}]%
        {levin1998all}
\bibfield{author}{\bibinfo{person}{Irwin~P Levin}, \bibinfo{person}{Sandra~L
  Schneider}, {and} \bibinfo{person}{Gary~J Gaeth}.}
  \bibinfo{year}{1998}\natexlab{}.
\newblock \showarticletitle{All frames are not created equal: A typology and
  critical analysis of framing effects}.
\newblock \bibinfo{journal}{\emph{Organizational behavior and human decision
  processes}} \bibinfo{volume}{76}, \bibinfo{number}{2} (\bibinfo{year}{1998}),
  \bibinfo{pages}{149--188}.
\newblock
\urldef\tempurl%
\url{https://doi.org/10.1006/obhd.1998.2804}
\showDOI{\tempurl}


\bibitem[\protect\citeauthoryear{Levin, Schnittjer, and Thee}{Levin
  et~al\mbox{.}}{1988}]%
        {levin1988information}
\bibfield{author}{\bibinfo{person}{Irwin~P Levin}, \bibinfo{person}{Sara~K
  Schnittjer}, {and} \bibinfo{person}{Shannon~L Thee}.}
  \bibinfo{year}{1988}\natexlab{}.
\newblock \showarticletitle{Information framing effects in social and personal
  decisions}.
\newblock \bibinfo{journal}{\emph{Journal of Experimental Social Psychology}}
  \bibinfo{volume}{24}, \bibinfo{number}{6} (\bibinfo{year}{1988}),
  \bibinfo{pages}{520--529}.
\newblock
\urldef\tempurl%
\url{https://doi.org/10.1016/0022-1031(88)90050-9}
\showDOI{\tempurl}


\bibitem[\protect\citeauthoryear{Li, Li, Wong, Feng, and Tan}{Li
  et~al\mbox{.}}{2005}]%
        {discriminative-pattern-mining}
\bibfield{author}{\bibinfo{person}{Haiquan Li}, \bibinfo{person}{Jinyan Li},
  \bibinfo{person}{Limsoon Wong}, \bibinfo{person}{Mengling Feng}, {and}
  \bibinfo{person}{{Yap Peng} Tan}.} \bibinfo{year}{2005}\natexlab{}.
\newblock \showarticletitle{Relative risk and odds ratio: A data mining
  perspective}. \bibinfo{pages}{368--377}.
\newblock


\bibitem[\protect\citeauthoryear{Lim, Dawson, Joksimovic, and
  Ga\v{s}evi\'{c}}{Lim et~al\mbox{.}}{2019}]%
        {Lim2019}
\bibfield{author}{\bibinfo{person}{Lisa Lim}, \bibinfo{person}{Shane Dawson},
  \bibinfo{person}{Srecko Joksimovic}, {and} \bibinfo{person}{Dragan
  Ga\v{s}evi\'{c}}.} \bibinfo{year}{2019}\natexlab{}.
\newblock \showarticletitle{Exploring Students' Sensemaking of Learning
  Analytics Dashboards: Does Frame of Reference Make a Difference?}. In
  \bibinfo{booktitle}{\emph{Proceedings of the 9th International Conference on
  Learning Analytics \& Knowledge}} (Tempe, AZ, USA)
  \emph{(\bibinfo{series}{LAK19})}. \bibinfo{publisher}{Association for
  Computing Machinery}, \bibinfo{address}{New York, NY, USA},
  \bibinfo{pages}{250–259}.
\newblock
\showISBNx{9781450362566}
\urldef\tempurl%
\url{https://doi.org/10.1145/3303772.3303804}
\showDOI{\tempurl}


\bibitem[\protect\citeauthoryear{Long and Siemens}{Long and Siemens}{2014}]%
        {LongSieme2014hy}
\bibfield{author}{\bibinfo{person}{Phillip Long} {and} \bibinfo{person}{George
  Siemens}.} \bibinfo{year}{2014}\natexlab{}.
\newblock \showarticletitle{Penetrating the fog: analytics in learning and
  education}.
\newblock \bibinfo{journal}{\emph{Italian Journal of Educational Technology}}
  \bibinfo{volume}{22}, \bibinfo{number}{3} (\bibinfo{date}{December}
  \bibinfo{year}{2014}), \bibinfo{pages}{132--137}.
\newblock
\showISSN{2532-4632}
\urldef\tempurl%
\url{https://www.learntechlib.org/p/183382}
\showURL{%
\tempurl}


\bibitem[\protect\citeauthoryear{Lundberg and Lee}{Lundberg and Lee}{2017}]%
        {NIPS2017_7062}
\bibfield{author}{\bibinfo{person}{Scott~M. Lundberg} {and}
  \bibinfo{person}{Su-In Lee}.} \bibinfo{year}{2017}\natexlab{}.
\newblock \showarticletitle{A Unified Approach to Interpreting Model
  Predictions}. In \bibinfo{booktitle}{\emph{Proceedings of the 31st
  International Conference on Neural Information Processing Systems}}
  \emph{(\bibinfo{series}{NIPS'17})}. \bibinfo{publisher}{Curran Associates
  Inc.}, \bibinfo{address}{Red Hook, NY, USA}, \bibinfo{pages}{4768–4777}.
\newblock
\showISBNx{9781510860964}
\urldef\tempurl%
\url{https://doi.org/10.5555/3295222.3295230}
\showDOI{\tempurl}


\bibitem[\protect\citeauthoryear{Maclellan}{Maclellan}{2004}]%
        {Maclellan2004}
\bibfield{author}{\bibinfo{person}{Effie Maclellan}.}
  \bibinfo{year}{2004}\natexlab{}.
\newblock \showarticletitle{How convincing is alternative assessment for use in
  higher education?}
\newblock \bibinfo{journal}{\emph{Assessment \& Evaluation in Higher
  Education}} \bibinfo{volume}{29}, \bibinfo{number}{3} (\bibinfo{year}{2004}),
  \bibinfo{pages}{311--321}.
\newblock
\urldef\tempurl%
\url{https://doi.org/10.1080/0260293042000188267}
\showDOI{\tempurl}


\bibitem[\protect\citeauthoryear{Main and Ost}{Main and Ost}{2014}]%
        {Main2014ImpactOfLetterGrades}
\bibfield{author}{\bibinfo{person}{Joyce~B. Main} {and} \bibinfo{person}{Ben
  Ost}.} \bibinfo{year}{2014}\natexlab{}.
\newblock \showarticletitle{The Impact of Letter Grades on Student Effort,
  Course Selection, and Major Choice: A Regression-Discontinuity Analysis}.
\newblock \bibinfo{journal}{\emph{The Journal of Economic Education}}
  \bibinfo{volume}{45}, \bibinfo{number}{1} (\bibinfo{year}{2014}),
  \bibinfo{pages}{1--10}.
\newblock
\urldef\tempurl%
\url{https://doi.org/10.1080/00220485.2014.859953}
\showDOI{\tempurl}


\bibitem[\protect\citeauthoryear{McElroy and Seta}{McElroy and Seta}{2003}]%
        {mcelroy2003framing}
\bibfield{author}{\bibinfo{person}{Todd McElroy} {and} \bibinfo{person}{John~J
  Seta}.} \bibinfo{year}{2003}\natexlab{}.
\newblock \showarticletitle{Framing effects: An analytic--holistic
  perspective}.
\newblock \bibinfo{journal}{\emph{Journal of Experimental Social Psychology}}
  \bibinfo{volume}{39}, \bibinfo{number}{6} (\bibinfo{year}{2003}),
  \bibinfo{pages}{610--617}.
\newblock
\urldef\tempurl%
\url{https://doi.org/10.1016/S0022-1031(03)00036-2}
\showDOI{\tempurl}


\bibitem[\protect\citeauthoryear{M\'{e}ndez, Hinrichs, and Nacenta}{M\'{e}ndez
  et~al\mbox{.}}{2017}]%
        {Mendez2017}
\bibfield{author}{\bibinfo{person}{Gonzalo~Gabriel M\'{e}ndez},
  \bibinfo{person}{Uta Hinrichs}, {and} \bibinfo{person}{Miguel~A. Nacenta}.}
  \bibinfo{year}{2017}\natexlab{}.
\newblock \showarticletitle{Bottom-up vs. Top-down: Trade-Offs in Efficiency,
  Understanding, Freedom and Creativity with InfoVis Tools}. In
  \bibinfo{booktitle}{\emph{Proceedings of the 2017 CHI Conference on Human
  Factors in Computing Systems}} (Denver, Colorado, USA)
  \emph{(\bibinfo{series}{CHI '17})}. \bibinfo{publisher}{Association for
  Computing Machinery}, \bibinfo{address}{New York, NY, USA},
  \bibinfo{pages}{841–852}.
\newblock
\showISBNx{9781450346559}
\urldef\tempurl%
\url{https://doi.org/10.1145/3025453.3025942}
\showDOI{\tempurl}


\bibitem[\protect\citeauthoryear{M{\'e}ndez, Nacenta, and
  Vandenheste}{M{\'e}ndez et~al\mbox{.}}{2016}]%
        {Mendez2016}
\bibfield{author}{\bibinfo{person}{Gonzalo~Gabriel M{\'e}ndez},
  \bibinfo{person}{Miguel~A. Nacenta}, {and} \bibinfo{person}{Sebastien
  Vandenheste}.} \bibinfo{year}{2016}\natexlab{}.
\newblock \showarticletitle{iVoLVER: Interactive Visual Language for
  Visualization Extraction and Reconstruction}. In
  \bibinfo{booktitle}{\emph{Proceedings of the 2016 CHI Conference on Human
  Factors in Computing Systems}} (Santa Clara, California, USA)
  \emph{(\bibinfo{series}{CHI '16})}. \bibinfo{publisher}{ACM},
  \bibinfo{address}{New York, NY, USA}, \bibinfo{pages}{4073--4085}.
\newblock
\showISBNx{978-1-4503-3362-7}
\urldef\tempurl%
\url{https://doi.org/10.1145/2858036.2858435}
\showDOI{\tempurl}


\bibitem[\protect\citeauthoryear{Munzner}{Munzner}{2014}]%
        {MunznerBook}
\bibfield{author}{\bibinfo{person}{Tamara Munzner}.}
  \bibinfo{year}{2014}\natexlab{}.
\newblock \bibinfo{booktitle}{\emph{Visualization {Analysis} and {Design}}
  (\bibinfo{edition}{har/psc edition} ed.)}.
\newblock \bibinfo{publisher}{A K Peters/CRC Press}, \bibinfo{address}{Boca
  Raton}.
\newblock
\showISBNx{9781466508910}
\urldef\tempurl%
\url{http://www.cs.ubc.ca/~tmm/vadbook}
\showURL{%
\tempurl}


\bibitem[\protect\citeauthoryear{Nelson and Oxley}{Nelson and Oxley}{1999}]%
        {nelson1999issue}
\bibfield{author}{\bibinfo{person}{Thomas~E Nelson} {and}
  \bibinfo{person}{Zoe~M Oxley}.} \bibinfo{year}{1999}\natexlab{}.
\newblock \showarticletitle{Issue framing effects on belief importance and
  opinion}.
\newblock \bibinfo{journal}{\emph{The journal of Politics}}
  \bibinfo{volume}{61}, \bibinfo{number}{4} (\bibinfo{year}{1999}),
  \bibinfo{pages}{1040--1067}.
\newblock
\urldef\tempurl%
\url{https://doi.org/10.2307/2647553}
\showDOI{\tempurl}


\bibitem[\protect\citeauthoryear{Ognjanovic, Gasevic, and Dawson}{Ognjanovic
  et~al\mbox{.}}{2016}]%
        {Ognjanovic2016}
\bibfield{author}{\bibinfo{person}{Ivana Ognjanovic}, \bibinfo{person}{Dragan
  Gasevic}, {and} \bibinfo{person}{Shane Dawson}.}
  \bibinfo{year}{2016}\natexlab{}.
\newblock \showarticletitle{Using institutional data to predict student course
  selections in higher education}.
\newblock \bibinfo{journal}{\emph{The Internet and Higher Education}}
  \bibinfo{volume}{29} (\bibinfo{year}{2016}), \bibinfo{pages}{49 -- 62}.
\newblock
\showISSN{1096-7516}
\urldef\tempurl%
\url{https://doi.org/10.1016/j.iheduc.2015.12.002}
\showDOI{\tempurl}


\bibitem[\protect\citeauthoryear{Oh, Slovacek, Tucker, and Hafner}{Oh
  et~al\mbox{.}}{2003}]%
        {oh2003assessment}
\bibfield{author}{\bibinfo{person}{Deborah~M. Oh}, \bibinfo{person}{Simeon
  Slovacek}, \bibinfo{person}{Susan Tucker}, {and} \bibinfo{person}{Ann
  Hafner}.} \bibinfo{year}{2003}\natexlab{}.
\newblock \showarticletitle{Assessment Outcomes of Pre-service Teachers}.
\newblock \bibinfo{journal}{\emph{Assessment \& Evaluation in Higher
  Education}} \bibinfo{volume}{28}, \bibinfo{number}{3} (\bibinfo{year}{2003}),
  \bibinfo{pages}{279--295}.
\newblock
\urldef\tempurl%
\url{https://doi.org/10.1080/0260293032000059630}
\showDOI{\tempurl}


\bibitem[\protect\citeauthoryear{Pham, Virlet, Lavenier, and Termier}{Pham
  et~al\mbox{.}}{2019}]%
        {Pham2019}
\bibfield{author}{\bibinfo{person}{Hoang~Son Pham}, \bibinfo{person}{Gwendal
  Virlet}, \bibinfo{person}{Dominique Lavenier}, {and}
  \bibinfo{person}{Alexandre Termier}.} \bibinfo{year}{2019}\natexlab{}.
\newblock \showarticletitle{{Statistically Significant Discriminative Patterns
  Searching}}.
\newblock \bibinfo{journal}{\emph{Lecture Notes in Computer Science}}
  (\bibinfo{year}{2019}), \bibinfo{pages}{105--115}.
\newblock
\showISBNx{9783030275204}
\showISSN{1611-3349}
\urldef\tempurl%
\url{https://doi.org/10.1007/978-3-030-27520-4_8}
\showDOI{\tempurl}


\bibitem[\protect\citeauthoryear{Rotherham and Willingham}{Rotherham and
  Willingham}{2010}]%
        {rotherham201021st}
\bibfield{author}{\bibinfo{person}{Andrew~J Rotherham} {and}
  \bibinfo{person}{Daniel~T Willingham}.} \bibinfo{year}{2010}\natexlab{}.
\newblock \showarticletitle{''21st-century'' skills}.
\newblock \bibinfo{journal}{\emph{American Educator}} \bibinfo{volume}{17},
  \bibinfo{number}{1} (\bibinfo{year}{2010}), \bibinfo{pages}{17--20}.
\newblock


\bibitem[\protect\citeauthoryear{Ruiz, Charleer, Urretavizcaya, Klerkx,
  Fern\'{a}ndez-Castro, and Duval}{Ruiz et~al\mbox{.}}{2016}]%
        {10.1145/2883851.2883888}
\bibfield{author}{\bibinfo{person}{Samara Ruiz}, \bibinfo{person}{Sven
  Charleer}, \bibinfo{person}{Maite Urretavizcaya}, \bibinfo{person}{Joris
  Klerkx}, \bibinfo{person}{Isabel Fern\'{a}ndez-Castro}, {and}
  \bibinfo{person}{Erik Duval}.} \bibinfo{year}{2016}\natexlab{}.
\newblock \showarticletitle{{Supporting Learning by Considering Emotions:
  Tracking and Visualization a Case Study}}. In
  \bibinfo{booktitle}{\emph{Proceedings of the Sixth International Conference
  on Learning Analytics \& Knowledge}}. \bibinfo{pages}{254–263}.
\newblock
\showISBNx{9781450341905}
\urldef\tempurl%
\url{https://doi.org/10.1145/2883851.2883888}
\showDOI{\tempurl}


\bibitem[\protect\citeauthoryear{Ryan and Deci}{Ryan and Deci}{2000a}]%
        {Ryan2000WhenRewardsCompeteWithNature}
\bibfield{author}{\bibinfo{person}{Richard~M. Ryan} {and}
  \bibinfo{person}{Edward~L. Deci}.} \bibinfo{year}{2000}\natexlab{a}.
\newblock \showarticletitle{Chapter 2 - When rewards compete with nature: The
  undermining of intrinsic motivation and Self-Regulation}.
\newblock In \bibinfo{booktitle}{\emph{Intrinsic and Extrinsic Motivation}},
  \bibfield{editor}{\bibinfo{person}{Carol Sansone} {and}
  \bibinfo{person}{Judith~M. Harackiewicz}} (Eds.).
  \bibinfo{publisher}{Academic Press}, \bibinfo{address}{San Diego},
  \bibinfo{pages}{13 -- 54}.
\newblock
\showISSN{18716148}
\urldef\tempurl%
\url{https://doi.org/10.1016/B978-012619070-0/50024-6}
\showDOI{\tempurl}


\bibitem[\protect\citeauthoryear{Ryan and Deci}{Ryan and Deci}{2000b}]%
        {Ryan2000SelfDetermination}
\bibfield{author}{\bibinfo{person}{Richard~M Ryan} {and}
  \bibinfo{person}{Edward~L Deci}.} \bibinfo{year}{2000}\natexlab{b}.
\newblock \showarticletitle{Self-determination theory and the facilitation of
  intrinsic motivation, social development, and well-being.}
\newblock \bibinfo{journal}{\emph{American psychologist}} \bibinfo{volume}{55},
  \bibinfo{number}{1} (\bibinfo{year}{2000}), \bibinfo{pages}{68}.
\newblock
\urldef\tempurl%
\url{https://doi.org/10.1037/0003-066X.55.1.68}
\showDOI{\tempurl}


\bibitem[\protect\citeauthoryear{Ryoo and Linn}{Ryoo and Linn}{2014}]%
        {ryoo2014designing}
\bibfield{author}{\bibinfo{person}{Kihyun Ryoo} {and} \bibinfo{person}{Marcia~C
  Linn}.} \bibinfo{year}{2014}\natexlab{}.
\newblock \showarticletitle{Designing guidance for interpreting dynamic
  visualizations: Generating versus reading explanations}.
\newblock \bibinfo{journal}{\emph{Journal of Research in Science Teaching}}
  \bibinfo{volume}{51}, \bibinfo{number}{2} (\bibinfo{year}{2014}),
  \bibinfo{pages}{147--174}.
\newblock
\urldef\tempurl%
\url{https://doi.org/10.1002/tea.21128}
\showDOI{\tempurl}


\bibitem[\protect\citeauthoryear{Sample}{Sample}{2018}]%
        {sample2018implications}
\bibfield{author}{\bibinfo{person}{Cassandra Sample}.}
  \bibinfo{year}{2018}\natexlab{}.
\newblock \showarticletitle{Implications of Elective Course Selection by Grade
  Maximization}.
\newblock  (\bibinfo{year}{2018}).
\newblock


\bibitem[\protect\citeauthoryear{Schwartz, Burgett, Blue, Donnelly, and
  Sloan}{Schwartz et~al\mbox{.}}{1997}]%
        {schwartz1997problem}
\bibfield{author}{\bibinfo{person}{Richard~W Schwartz},
  \bibinfo{person}{James~E Burgett}, \bibinfo{person}{Amy~V Blue},
  \bibinfo{person}{Michael~B Donnelly}, {and} \bibinfo{person}{David~A Sloan}.}
  \bibinfo{year}{1997}\natexlab{}.
\newblock \showarticletitle{Problem-based learning and performance-based
  testing: Effective alternatives for undergraduate surgical education and
  assessment of student performance}.
\newblock \bibinfo{journal}{\emph{Medical Teacher}} \bibinfo{volume}{19},
  \bibinfo{number}{1} (\bibinfo{year}{1997}), \bibinfo{pages}{19--23}.
\newblock
\urldef\tempurl%
\url{https://doi.org/10.3109/01421599709019341}
\showDOI{\tempurl}


\bibitem[\protect\citeauthoryear{Smith}{Smith}{1995}]%
        {Smith1995UnintendedConsequences}
\bibfield{author}{\bibinfo{person}{Peter Smith}.}
  \bibinfo{year}{1995}\natexlab{}.
\newblock \showarticletitle{On the unintended consequences of publishing
  performance data in the public sector}.
\newblock \bibinfo{journal}{\emph{International Journal of Public
  Administration}} \bibinfo{volume}{18}, \bibinfo{number}{2-3}
  (\bibinfo{year}{1995}), \bibinfo{pages}{277--310}.
\newblock
\urldef\tempurl%
\url{https://doi.org/10.1080/01900699508525011}
\showDOI{\tempurl}


\bibitem[\protect\citeauthoryear{Struyven, Dochy, Janssens, Schelfhou, and
  Gielen}{Struyven et~al\mbox{.}}{2006}]%
        {Struyven2006}
\bibfield{author}{\bibinfo{person}{Katrien Struyven}, \bibinfo{person}{Filip
  Dochy}, \bibinfo{person}{Steven Janssens}, \bibinfo{person}{Wouter
  Schelfhou}, {and} \bibinfo{person}{Sarah Gielen}.}
  \bibinfo{year}{2006}\natexlab{}.
\newblock \showarticletitle{The overall effects of end-of-course assessment on
  student performance: A comparison between multiple choice testing, peer
  assessment, case-based assessment and portfolio assessment}.
\newblock \bibinfo{journal}{\emph{Studies in Educational Evaluation}}
  \bibinfo{volume}{32}, \bibinfo{number}{3} (\bibinfo{year}{2006}),
  \bibinfo{pages}{202 -- 222}.
\newblock
\showISSN{0191-491X}
\urldef\tempurl%
\url{https://doi.org/10.1016/j.stueduc.2006.08.002}
\showDOI{\tempurl}


\bibitem[\protect\citeauthoryear{Ting and Lee}{Ting and Lee}{2012}]%
        {ting2012understanding}
\bibfield{author}{\bibinfo{person}{Ding~Hooi Ting} {and}
  \bibinfo{person}{Christina Kwai~Choi Lee}.} \bibinfo{year}{2012}\natexlab{}.
\newblock \showarticletitle{Understanding students’ choice of electives and
  its implications}.
\newblock \bibinfo{journal}{\emph{Studies in Higher Education}}
  \bibinfo{volume}{37}, \bibinfo{number}{3} (\bibinfo{year}{2012}),
  \bibinfo{pages}{309--325}.
\newblock


\bibitem[\protect\citeauthoryear{Tourangeau, Couper, and Conrad}{Tourangeau
  et~al\mbox{.}}{2004}]%
        {Tourangeau2004SpacingPositionOrder}
\bibfield{author}{\bibinfo{person}{Roger Tourangeau}, \bibinfo{person}{Mick~P.
  Couper}, {and} \bibinfo{person}{Frederick Conrad}.}
  \bibinfo{year}{2004}\natexlab{}.
\newblock \showarticletitle{{Spacing, Position, and Order: Interpretive
  Heuristics for Visual Features of Survey Questions}}.
\newblock \bibinfo{journal}{\emph{Public Opinion Quarterly}}
  \bibinfo{volume}{68}, \bibinfo{number}{3} (\bibinfo{date}{09}
  \bibinfo{year}{2004}), \bibinfo{pages}{368--393}.
\newblock
\showISSN{0033-362X}
\urldef\tempurl%
\url{https://doi.org/10.1093/poq/nfh035}
\showDOI{\tempurl}


\bibitem[\protect\citeauthoryear{Vieira, Parsons, and Byrd}{Vieira
  et~al\mbox{.}}{2018}]%
        {Viera2018_VLA}
\bibfield{author}{\bibinfo{person}{Camilo Vieira}, \bibinfo{person}{Paul
  Parsons}, {and} \bibinfo{person}{Vetria Byrd}.}
  \bibinfo{year}{2018}\natexlab{}.
\newblock \showarticletitle{{Visual Learning Analytics of Educational Data: A
  Systematic Literature Review and Research Agenda}}.
\newblock \bibinfo{journal}{\emph{Computers \& Education}}
  \bibinfo{volume}{122} (\bibinfo{year}{2018}), \bibinfo{pages}{119--135}.
\newblock
\urldef\tempurl%
\url{https://doi.org/10.1016/j.compedu.2018.03.018}
\showDOI{\tempurl}


\bibitem[\protect\citeauthoryear{Walny, Huron, and Carpendale}{Walny
  et~al\mbox{.}}{2015}]%
        {walny2015Sketching}
\bibfield{author}{\bibinfo{person}{Jagoda Walny}, \bibinfo{person}{Samuel
  Huron}, {and} \bibinfo{person}{Sheelagh Carpendale}.}
  \bibinfo{year}{2015}\natexlab{}.
\newblock \showarticletitle{An Exploratory Study of Data Sketching for Visual
  Representation}.
\newblock \bibinfo{journal}{\emph{Computer Graphics Forum}}
  \bibinfo{volume}{34}, \bibinfo{number}{3} (\bibinfo{year}{2015}),
  \bibinfo{pages}{231--240}.
\newblock
\showISSN{1467-8659}
\urldef\tempurl%
\url{https://doi.org/10.1111/cgf.12635}
\showDOI{\tempurl}


\bibitem[\protect\citeauthoryear{Ware}{Ware}{2012}]%
        {WareBook}
\bibfield{author}{\bibinfo{person}{Colin Ware}.}
  \bibinfo{year}{2012}\natexlab{}.
\newblock \bibinfo{booktitle}{\emph{Information {Visualization}}
  (\bibinfo{edition}{3rd revised edition edition} ed.)}.
\newblock \bibinfo{publisher}{Morgan Kaufmann}, \bibinfo{address}{Waltham, MA}.
\newblock
\showISBNx{9780123814647}


\bibitem[\protect\citeauthoryear{Zhou and Yu}{Zhou and Yu}{2008}]%
        {4722942}
\bibfield{author}{\bibinfo{person}{Qing Zhou} {and} \bibinfo{person}{Fang Yu}.}
  \bibinfo{year}{2008}\natexlab{}.
\newblock \showarticletitle{{Knowledge-Based Major Choosing Decision Making for
  Remote Students}}. In \bibinfo{booktitle}{\emph{International Conference on
  Computer Science and Software Engineering}}, Vol.~\bibinfo{volume}{5}.
  \bibinfo{pages}{474--478}.
\newblock
\urldef\tempurl%
\url{https://doi.org/10.1109/CSSE.2008.379}
\showDOI{\tempurl}


\end{thebibliography}

\end{document}